\documentclass[draft]{agujournal2019}
\usepackage{url} %this package should fix any errors with URLs in refs.
\usepackage[inline]{trackchanges} %for better track changes. finalnew option will compile document with changes incorporated.
\usepackage{soul}
\usepackage{longtable}
\usepackage{float}

\usepackage{amsfonts}
\usepackage{amsmath}
\usepackage{amssymb}
\usepackage{mathrsfs}
\usepackage{natbib}
\usepackage{rotating}
\usepackage{hyperref}
\draftfalse

\journalname{JGR: Planets}

\begin{document}

%%%%%%%%%%%%%%%%%%%%%%%%%%%%%%%%%%%%%%%%%%%%%%%
%  TITLE
%
% (A title should be specific, informative, and brief. Use
% abbreviations only if they are defined in the abstract. Titles that
% start with general keywords then specific terms are optimized in
% searches)
%
%%%%%%%%%%%%%%%%%%%%%%%%%%%%%%%%%%%%%%%%%%%%%%%

% Example: \title{This is a test title}

\title{On the Efficacy of Ocean Formation with a Primordial Hydrogen Atmosphere}

%%%%%%%%%%%%%%%%%%%%%%%%%%%%%%%%%%%%%%%%%%%%%%%
%
%  AUTHORS AND AFFILIATIONS
%
%%%%%%%%%%%%%%%%%%%%%%%%%%%%%%%%%%%%%%%%%%%%%%%

% Authors are individuals who have significantly contributed to the
% research and preparation of the article. Group authors are allowed, if
% each author in the group is separately identified in an appendix.)

% List authors by first name or initial followed by last name and
% separated by commas. Use \affil{} to number affiliations, and
% \thanks{} for author notes.
% Additional author notes should be indicated with \thanks{} (for
% example, for current addresses).

% Example: \authors{A. B. Author\affil{1}\thanks{Current address, Antartica}, B. C. Author\affil{2,3}, and D. E.
% Author\affil{3,4}\thanks{Also funded by Monsanto.}}

\authors{Darius Modirrousta-Galian\affil{1,2,3}, Jun Korenaga\affil{3}}
\affiliation{1}{Tsung-Dao Lee Institute, Shanghai Jiao Tong University, 1 Lisuo Road, Shanghai, 201210, China}
\affiliation{2}{School of Physics and Astronomy, Shanghai Jiao Tong University, 800 Dongchuan Road, Shanghai, 200240, China}
\affiliation{3}{Yale University, Department of Earth and Planetary Sciences, 210 Whitney Ave., New Haven, CT 06511, USA}

\correspondingauthor{Darius Modirrousta-Galian}{darius.modirrousta-galian@yale.edu}

%%%%%%%%%%%%%%%%%%%%%%%%%%%%%%%%%%%%%%%%%%%%%%%
% KEY POINTS
%%%%%%%%%%%%%%%%%%%%%%%%%%%%%%%%%%%%%%%%%%%%%%%
\begin{keypoints}
\item	Convective mixing efficiency in a magma ocean is evaluated with the theory of thermochemical convection.
\item	Compositional buoyancy from iron loss can halt magma ocean mixing.
\item	Hydrogen dissolution likely plays a minor role in Earth's ocean formation.
\end{keypoints}

\begin{abstract}
It has been suggested that Earth's present water budget formed from oxidation reactions between its initial hydrogen-rich primordial atmosphere and its magma ocean. Here we examine this hypothesis by building a comprehensive atmosphere-magma ocean model. We find that water formation is unlikely for two reasons. First, any water formed from oxidation reactions in the magma ocean would quickly outgass because of the water-poor atmosphere above. Second, the top boundary layer of the magma ocean becomes stable against convection because the oxidation reactions produce metallic iron, which sinks to the core of a growing Earth. This iron loss makes the top boundary layer significantly more buoyant than the rest of the magma, thus becoming stable against mixing. Our results suggest that hydrogen dissolution is unlikely to play a major role in the formation of Earth's oceans.
\end{abstract}

\section*{Plain Language Summary}
A growing Earth was hot, likely covered with magma oceans, and surrounded by a thick hydrogen-rich atmosphere. It has been suggested that Earth's present-day oceans might have formed when this hydrogen mixed with the magma on the surface. We developed a numerical model to test this idea and identified two key problems. First, any water that formed would quickly escape from the magma, turn to steam, and be absorbed by the atmosphere because it was dry and acted as a sponge. Second, the chemical reactions involved in this process create metallic iron that sinks deep into Earth, leaving behind a lighter magma layer on top. This lighter iron-poor layer acted like a lid, preventing further hydrogen from mixing with the magma. These issues prevent water from accumulating in the magma and stop the process from continuing. Our findings suggest that Earth's oceans are unlikely to have formed through this mechanism.

\section{Introduction}
\label{sec:intro}

The origin of water on Earth has long been debated because, in the standard model of planetary formation, volatiles condense beyond the ice line during the protoplanetary disk phase, which is thought to be located at approximately $3~{\rm AU}$ from the Sun \citep{Hayashi1981,Morbidelli2000,Podolak2004}. Some authors suggest that turbulence within the protoplanetary disk may have brought the ice line to orbital distances closer than $1~{\rm AU}$, providing a possible pathway for a growing Earth to acquire water \citep{Davis2005,Garaud2007,Oka2011}. Others suggest instead that dead zones within the protoplanetary disk, that is, regions of low turbulence, greatly reduce the flow of icy material to such close distances so that the effective ice line is still located at ${\sim}3~{\rm AU}$ \citep{Martin2012,Martin2013}. It has also been proposed that Jupiter's formation and migration may have allowed for significant material mixing within the protoplanetary disk, delivering $10{–}100$ Earth oceans worth of water in the form of chondrites to the inner solar system \citep{Raymond2004,Marov2018,Ogihara2023}. The orbital dynamics of planetesimals are, however, inherently chaotic so that they exhibit a sensitive dependence on initial conditions \citep{Wisdom1980,Heggie1991,Hut2002}. Whereas deterministic integration methods can quantify the likelihood of various outcomes for any given set of inputs, our limited knowledge of the true initial conditions introduces an inescapable uncertainty. Thus, even if the dynamics themselves are well understood, the overall probability of achieving the necessary conditions for substantial ice accretion remains uncertain. Another possibility is that Earth's building blocks were already water-rich \citep{Piani2020}. However, this hypothesis also faces challenges because Earth's poorly constrained accretion history and the broad range of permissible geochemical compositions make it difficult to verify \citep{Mezger2020}.

An alternative hypothesis, which has gained popularity in recent years, posits that Earth's oceans formed from oxidation reactions between the initial hydrogen-rich primordial atmosphere and the magma ocean of a growing Earth. \citet{Sasaki1990} was the first to investigate hydrogen dissolution in magma oceans and its influence on the bulk properties of a growing Earth. He found that a hydrogen-rich atmosphere would react with oxides in magma to form water, reducing the magma ocean and oxidizing the atmosphere. These arguments were then generalized to exoplanets to show that water oceans may be ubiquitous in the universe \citep{Ikoma2006,Seo2024}. \citet{Chachan2018} and \cite{Kite2019} quantified the amount of hydrogen dissolution in magma oceans under a hydrogen-rich atmosphere, and \citet{Kite2020} and \citet{Kite2021} evaluated its impact on magma ocean chemistry. \citet{Wu2018} and \citet{Young2023} developed more comprehensive models to explain the origin of Earth's oceans, with markedly different conclusions. \citet{Wu2018} found that Earth accreted approximately seven oceans worth of water from chondritic materials and ingassed only ${\sim}0.06$ ocean from its primordial hydrogen-rich atmosphere. In contrast, \citet{Young2023} argue that Earth's oceans, core density, and mantle redox state all arose from the chemical equilibrium between Earth's initial hydrogen-rich atmosphere and magma ocean. We note that both possibilities are consistent with the boundary layer analysis of \citet{Olson2018}, which suggests that between $0.1{-}4$ Earth oceans can form from hydrogen ingassing.

In light of the diverse and sometimes conflicting hypotheses about the origin of water on Earth, it is important to quantify the efficiency of atmospheric hydrogen mixing in a growing Earth's magma ocean and determine its contribution to the origin of water on our planet. In this study, we approach this problem by evaluating the efficiency of convective mixing in a magma ocean and its contribution to ocean formation. This paper is structured as follows: Section~\ref{sec:nomenclature} defines key terms; Sections~\ref{sec:protoplanetary}, \ref{sec:atm_model}, and \ref{sec:magma_model} detail our protoplanetary disk, atmospheric, and magma ocean models, respectively; Section~\ref{sec:interactions} outlines our approach to modeling fluid dynamics and chemistry for atmosphere-interior interactions; Section~\ref{sec:mechanical mixing} explores the effects of mixing from impacts and wind waves; and Section~\ref{sec:results} presents and discusses our findings, and compares them with existing literature. This paper concludes with a summary of our ﬁndings.

\section{Nomenclature}
\label{sec:nomenclature}

We define a proto-Earth as the gravitational assembly of matter prior to the Moon-forming giant impact. The exact size of a proto-Earth is not known, and we therefore characterize it as comprising at least 50\% of Earth's current mass. This criterion ensures that the impacting body responsible for the Moon formation is smaller than a proto-Earth. Isotopic evidence suggests that this impact occurred approximately $50$ million years after the dispersal of the protoplanetary disk \citep{Barboni2017,Thiemens2019}, and it is therefore not investigated in this study that focuses only on the interactions of a proto-Earth with the protoplanetary disk. The formation pathway leading to a proto-Earth is unimportant for our purposes, thus avoiding the debate surrounding the planetesimal vs. pebble accretion models \citep[e.g.,][]{Brugger2020,Raymond2022}. In this paper, the term nucleus refers to the central condensed section of a growing Earth that consists of the mantle and the metallic core and that accretes gas from the surrounding protoplanetary disk. All equations, terms, parameters, and values in this paper are expressed in the International System of Units (SI).

\section{Protoplanetary disk model}
\label{sec:protoplanetary}

Isotopic data suggests that the precursor material to the rocky planets in our solar system formed within a protoplanetary disk \citep{Schiller2018,Burkhardt2021}. This material, referred to as planetesimals henceforth, was likely partially molten because of radiogenic heating \citep{Hevey2006,Sahijpal2007}. In the absence of an atmosphere, cooling is efficient at temperatures above the rheological transition \citep[i.e., at potential temperature $T{\gtrsim}1610~{\rm K}$ in case of an Earth-like mantle composition;][]{Solomatov2015,Korenaga2023} because of turbulent convection. Without an atmosphere, the thermal interaction between the disk and the magma ocean results in the cooling of magma and the heating of local protoplanetary disk gas. In contrast, cooling is inefficient at temperatures below the rheological transition because magma behaves rheologically like a solid. These contrasting behaviors suggest that growing planetary embryos without atmospheres in a protoplanetary disk are likely to have temperatures close to the rheological transition. In the presence of a hydrogen-rich atmosphere, however, cooling can be significantly slowed by thermal blanketing. This insulating effect occurs if the planetesimal has sufficient mass to host an atmosphere \citep{Hayashi1979,Hubbard1980,Vazan2020,Kite2020}.

We approximate the protoplanetary disk as being chemically homogeneous with a bulk mean molecular weight of $2.3~{\rm amu}$ and an isentropic exponent of $\gamma{=}7/5$. Our equations for the profiles of the radial midplane volume density and temperature are similar to those of \citet{Rafikov2006b}, which are based on the minimum mass solar nebula, that is, the lower limit to the amount of gas needed to form the solar system planets \citep{Weidenschilling1977,Hayashi1981}:
\begin{equation}
    \rho_{\rm d}^{\leftrightarrow}\left(a\right) = 2.4{\times}10^{-6} S_{\rm d}\left(\frac{a}{\rm AU}\right)^{-\frac{11}{4}},
\label{eq:disk_density}
\end{equation}
\begin{equation}
    T_{\rm d}^{\leftrightarrow}\left(a\right) = 250 \left(\frac{a}{\rm AU}\right)^{-\frac{1}{2}},
\label{eq:disk_temperature}
\end{equation}
with the superscript $\leftrightarrow$ denoting the radial component along the midplane and $a$ being the distance from the star. Equation~\ref{eq:disk_temperature} assumes a faint young sun \citep{Christensen2021} with heating caused only by stellar radiation. Midplane temperatures may be initially very high because of the conversion of gravitational potential energy into heat \citep{Nakamoto1994,Hueso2005,Li2021}, but efficient cooling rapidly lowers these temperatures close to those predicted by Equation~\ref{eq:disk_temperature}, particularly in regions with $a{\gtrsim}1~{\rm AU}$, where turbulent dissipative heating is less significant \citep{Chiang1997,Dalessio1998,Rafikov2006a,Lesniak2011}. The vertical temperature profile is small and it is therefore approximated as being isothermal \citep{Dalessio1998,Dullemond2002,Mori2019}.

The parameter $S_{\rm d}$ in Equation~\ref{eq:disk_density} is a scaling factor, with $S_{\rm d}{=}1$ corresponding to the minimum mass solar nebula of approximately $0.01~M_{\odot}$. Planet formation is, however, not perfectly efficient, and several mechanisms result in mass loss from protoplanetary disks, so a solar nebula of mass $0.1{-}1~M_{\odot}$ is a more appropriate reference for modeling planet formation \citep{Boss2002,Lodato2004,Meru2010,Nixon2018}. Recent simulations support this idea by indicating that newly formed protoplanetary disks typically have masses of $0.1{-}1~M_{\odot}$ around Sun-like stars \citep{Schib2021}. This is further corroborated by the newly discovered protoplanetary disk IRAS 23077+6707 \citep[``Dracula's Chivito";][]{Monsch2024}, the largest known as of writing, with a predicted mass of $0.2~M_{\odot}$ \citep{Berghea2024}. In the context of the solar system, isotopic analyses of iron meteorites suggest that Jupiter's nucleus formed within the first 0.5 Myr after the formation of the Sun. Proto-Jupiter then grew to approximately fifty Earth masses within 2 Myr and completed its formation within $4{-}5$~Myr \citep{Kruijer2017,Kruijer2020}. Because protoplanetary disks have \textit{e}-folding times of $2{-}3$~Myr \citep{Mamajek2009,Williams2011}, the minimum mass solar nebula would have undergone considerable decay within the period required for Jupiter's formation, indicating a need for more massive disks. Observations and theory thus generally suggest $S_{\rm d}{>}1$. At the same time, it is also difficult for a protoplanetary disk to be very large (e.g., $S_{\rm d}{\sim}100$) and yet maintain its spherical symmetry (i.e., Equations~\ref{eq:disk_density} and \ref{eq:disk_temperature}) because self-gravity may become important, causing the disk to fragment and form spirals that affect overall dynamics and structure \citep{Safronov1960,Toomre1964,Lin1987,Gammie2001,Cossins2010}. In this paper, we therefore adopt the minimum value of $S_{\rm d}{=}1$ and the maximum value of $S_{\rm d}{=}100$, with the most likely value being around $S_{\rm d}{=}10$. 

We can estimate the optical thickness of the protoplanetary disk as follows. If we assume that the gravitational force of the Sun is much greater than that of the protoplanetary disk, the vertical component of gravity in the disk is given by
\begin{equation}
    g = -GM_{\odot}\frac{z}{\left(a^{2}+z^{2}\right)^{\frac{3}{2}}},
\end{equation}
where $G$ is the gravitational constant, $M_{\odot}$ is the solar mass, and $z$ is its vertical thickness. Combining with the equation for hydrostatic equilibrium,
\begin{equation}
    \frac{{\rm d}P}{{\rm d}z}=-GM_{\odot}\rho \frac{z}{\left(a^{2}+z^{2}\right)^{\frac{3}{2}}},
\end{equation}
and using the ideal gas equation,
\begin{equation}
    \frac{{\rm d}P}{{\rm d}z} = \frac{k_{\rm B}T}{\bar{\mu}}\frac{{\rm d}\rho}{{\rm d}z} + \frac{k_{\rm B} \rho}{\bar{\mu}}\frac{{\rm d}T}{{\rm d}z},
\end{equation}
yields,
\begin{equation}
    \frac{1}{\rho}\frac{{\rm d}\rho}{{\rm d}z} = -\frac{1}{T}\frac{{\rm d}T}{{\rm d}z} - \frac{GM_{\odot}\bar{\mu}}{k_{\rm B}T} \frac{z}{\left(a^{2}+z^{2}\right)^{\frac{3}{2}}},
\end{equation}
with $\bar{\mu}$ being the mean molecular mass and $k_{\rm B}$ is Boltzmann's constant. The temperature gradient in the z-axis is very small because the radial distance to the star changes little with $z$ so that ${\rm d}T/{\rm d}z{\approx}0$. We also approximate $a^{2}{+}z^{2}{\approx}a^{2}$ so that
\begin{equation}
    \frac{1}{\rho}\frac{{\rm d}\rho}{{\rm d}z} = - \frac{GM_{\odot}\bar{\mu}}{k_{\rm B}T a^{3}} z.
\label{eq:integration}
\end{equation}
We set the Bondi radius, which is the location at which gas attains escape velocity (i.e., $v_{\rm esc}{=}\sqrt{2GM_{\rm p}/r}{=}v_{\rm sound}{=}\sqrt{\gamma k_{\rm B}T/\bar{\mu}}$, with $\gamma$ being the isentropic exponent), as the boundary between the proto-Earth and the protoplanetary disk, 
\begin{equation}
    R_{\rm B} = \frac{2GM_{\rm p}\bar{\mu}}{\gamma k_{\rm B}T_{\rm B}},
\label{eq:R_B}
\end{equation}
where $M_{\rm p}$ is the mass of a proto-Earth, $T_{\rm B}$ (close to but not equal to $T_{\rm d}\left(a_{\rm p}\right)$; Section~\ref{sec:atm_model}) is the temperature at the Bondi radius, and $a_{\rm p}$ is semi-major axis distance from the Sun. The mass of the nucleus is much larger than the atmospheric mass so that $M_{\rm n}{\approx}M_{\rm p}$. Therefore, we use $M_{\rm n}$ henceforth and we do not consider the energy contribution from the gravitational accretion of the primordial atmosphere. We scale the radius of a proto-Earth with its mass using $R_{\rm n}/R_{\oplus}{=}\left(M_{\rm n}/M_{\oplus}\right)^{1/4}$ \citep{Zeng2016,Muller2024}. Integrating Equation~\ref{eq:integration} from the Bondi radius upward and substituting in Equation~\ref{eq:R_B},
\begin{equation}
    \rho_{\rm d}^{\updownarrow}\left(z\right) = \rho_{\rm B}\exp{\left\{-\frac{\gamma}{4}\frac{M_{\odot}}{M_{\rm n}} \left(\frac{R_{\rm B}}{a_{\rm p}}\right)^{3} \left[\left(\frac{z}{R_{\rm B}}\right)^{2}{-}1\right]\right\}},
\label{eq:density_2}
\end{equation}
where the superscript $\updownarrow$ denotes the vertical component above the midplane and $\rho_{\rm B}{\equiv}\rho^{\leftrightarrow}_{\rm d}\left(a_{\rm p}\right)$ (Equation~\ref{eq:disk_density}). Equations~\ref{eq:disk_density} and \ref{eq:density_2} represent, respectively, the radial midplane and vertical volume density components of the protoplanetary disk. 

The optical depth at the Bondi radius varies with the latitude of a proto-Earth. At the equator (assuming zero planetary axial tilt), it is determined by the radial density profile (Equation~\ref{eq:disk_density}), whereas at the poles, it is determined by the vertical density profile (Equation~\ref{eq:density_2}). Latitudes in between would require a combination of both. The optical depth in the radial direction is virtually infinite because of abundant interplanetary gas. We may therefore evaluate the minimum optical depth at the Bondi radius by considering only the vertical component corresponding to the poles:
\begin{equation}
\begin{split}
    \tau_{\rm B} &= \rho_{\rm B}\bar{\kappa}\int^{\infty}_{R_{\rm B}} \exp{\left\{-\frac{\gamma}{4}\frac{M_{\odot}}{M_{\rm n}} \left(\frac{R_{\rm B}}{a_{\rm p}}\right)^{3} \left[\left(\frac{z}{R_{\rm B}}\right)^{2}{-}1\right]\right\}}~{\rm d}r \\
    & \approx \left(\frac{\pi}{\gamma}\right)^{\frac{1}{2}} \left(\frac{M_{\rm n}}{M_{\odot}}\right)^{\frac{1}{2}} \left(\frac{a_{\rm p}}{R_{\rm B}}\right)^{\frac{1}{2}} \rho_{\rm B} \bar{\kappa} a_{\rm p},
\end{split}
\end{equation}
with $\bar{\kappa}$ being the mean opacity. Combining with Equations~\ref{eq:disk_density}, \ref{eq:disk_temperature}, and \ref{eq:R_B},
\begin{equation}
    \tau_{\rm B} \approx S_{\rm d}\left(\frac{\bar{\kappa}}{10^{-4}~{\rm m^{2}~kg^{-1}}}\right)\left(\frac{a_{\rm p}}{\rm AU}\right)^{-\frac{3}{2}},
\label{eq:tau_B}
\end{equation}
for a solar-mass star and a protoplanetary disk of mean molecular weight $\bar{\mu}{=}2.3~{\rm amu}$. For typical protoplanetary disk opacities of $\bar{\kappa}{\sim}1~{\rm m^{2}~kg^{-1}}$ \citep{Ali1992,Henning1996,Chachan2021}, gas is optically thick for any reasonable choice of parameters. We thus consider an optically thick protoplanetary disk henceforth, which is also consistent with observations \citep[e.g.,][]{Wood2002,Xin2023,Rilinger2023}

\section{Atmosphere model}
\label{sec:atm_model}

The surface of a proto-Earth was likely molten from frequent impacts with planetesimals \citep{Goldreich1973,Wetherill1989,Johansen2007} and thermal blanketing \citep{Hayashi1979,Matsui1986,Abe1997}. The optically thick protoplanetary disk blocks incoming stellar radiation so that the high internal heat flux of a growing Earth dominates atmospheric dynamics; we thus consider only the internal heat component. We begin with the assumption that the atmosphere of a proto-Earth was fully radiative (see~\ref{sec:radiative_diffusion}),
\begin{equation}
    \left.\frac{{\rm d}T}{{\rm d}r}\right|_{\rm rad}= -\frac{3 \bar{\kappa}\rho L}{64 \pi r^{2} \sigma T^{3}},
\label{eq:dT_dr_rad}
\end{equation}
where $T$ is temperature, $\bar{\kappa}$ is the mean opacity, $\rho$ is density, $L$ is internal luminosity, $r$ is radial distance, and $\sigma$ is the Stefan-Boltzmann constant. Combining Equation~\ref{eq:dT_dr_rad} with the equation for hydrostatic equilibrium,
\begin{equation}
    \frac{{\rm d}P}{{\rm d}r} = -\frac{GM_{\rm n}\rho}{r^{2}},
\label{eq:hydrostatic}
\end{equation}
yields,
\begin{equation}
    \left.\frac{{\rm d}T}{{\rm d}P}\right|_{\rm rad}= \frac{3\bar{\kappa}L}{64 \pi \sigma T^{3} G M_{\rm n}}.
\label{eq:dT_dP_rad}
\end{equation}
Integrating Equation~\ref{eq:dT_dP_rad} with the boundary conditions set by those of the Bondi radius,
\begin{equation}
    P = P_{\rm B} + \frac{16 \pi \sigma GM_{\rm n}}{3 \bar{\kappa} L}\left(T^{4}-T_{\rm B}^{4}\right)
\label{eq:P_solution}
\end{equation}
which can be combined with the ideal gas equation and solved as follows,
\begin{equation}
    T^{3} = \frac{3\bar{\kappa}L \rho k_{\rm B}}{16 \pi \sigma G M_{\rm n} \bar{\mu}} \frac{1-\frac{P_{\rm B}}{P}}{1-\left(\frac{T_{\rm B}}{T}\right)^{4}}.
\label{eq:T3_rad}
\end{equation}
By combining Equations~\ref{eq:dT_dr_rad} and \ref{eq:T3_rad}, we obtain
\begin{equation}
    \left.\frac{{\rm d}T}{{\rm d}r}\right|_{\rm rad}= -\frac{GM_{\rm n} \bar{\mu}}{4 k_{\rm B} r^{2}} \frac{1-\left(\frac{T_{\rm B}}{T}\right)^{4}}{1-\frac{P_{\rm B}}{P}},
\label{eq:dT_dr_rad_2}
\end{equation}
where the rightmost term is close to unity for $T{\gg}T_{\rm B}$ and $P{\gg}P_{\rm B}$ but it becomes undefined at the Bondi radius where $T{=}T_{\rm B}$ and $P{=}P_{\rm B}$. L'Hôpital's Rule applies to univariate functions but it can be extended to multivariate functions for specific cases \citep[e.g.,][]{Lawlor2020}. We can prove that a limit does not exist for Equation~\ref{eq:T3_rad} by restricting the domain to individual lines through $(T{,}P){\rightarrow}\left(T_{\rm B}{,}P_{\rm B}\right)$, e.g., for $P{\propto}T$, the limit is 4, whereas for $P{\propto}T^{4}$, the limit is unity. We can address the undefined behavior of the rightmost term by eliminating the density dependency in Equation~\ref{eq:T3_rad} using the ideal gas law. The resulting expression can then be simplified by combining it with Equation~\ref{eq:P_solution}, removing the pressure dependency and expressing it as a function of temperature. Inserting into Equation~\ref{eq:dT_dr_rad_2}, we thus obtain,
\begin{equation}
    \left.\frac{{\rm d}T}{{\rm d}r}\right|_{\rm rad}= -\frac{GM_{\rm n} \bar{\mu}}{4 k_{\rm B} r^{2}} \left[\left(\frac{T_{\rm B}}{T}\right)^{4}\left(\frac{3 \bar{\kappa} P_{\rm B} L}{16 \pi \sigma G M_{\rm n} T_{\rm B}^{4}} - 1\right) + 1\right].
\label{eq:dT_prefinal}
\end{equation}
The physical significance of the dimensionless $3 \bar{\kappa} P_{\rm B} L/\left(16 \pi \sigma G M_{\rm n} T_{\rm B}^{4}\right)$ factor can be understood by considering the density gradient of the atmosphere. We start with the pressure gradient and hydrostatic equilibrium,
\begin{equation}
    \frac{{\rm d}P}{{\rm d}r} = \frac{k_{\rm B}T}{\mu}\frac{{\rm d}\rho}{{\rm d}r} + \frac{k_{\rm B} \rho}{\mu}\frac{{\rm d}T}{{\rm d}r} = -\rho \frac{GM_{\rm n}}{r^{2}},
\end{equation}
which can be solved for the density gradient,
\begin{equation}
    \frac{{\rm d}\ln{\left(\rho\right)}}{{\rm d}\ln{\left(r\right)}} = -\frac{GM_{\rm n}\mu}{k_{\rm B} Tr} - \frac{{\rm d}\ln{\left(T\right)}}{{\rm d}\ln{\left(r\right)}},
\end{equation}
and combined with Equation~\ref{eq:dT_prefinal}
\begin{equation}
    \left.\frac{{\rm d}\ln{\left(\rho\right)}}{{\rm d}\ln{\left(r\right)}}\right|_{\rm rad}= \frac{GM_{\rm n} \bar{\mu}}{4 k_{\rm B}T r} \left[\left(\frac{T_{\rm B}}{T}\right)^{4}\left(\frac{3 \bar{\kappa} P_{\rm B} L}{16 \pi \sigma G M_{\rm n} T_{\rm B}^{4}} - 1\right) - 3\right].
\label{eq:drho_prefinal}
\end{equation}
The density gradient can never be greater than zero otherwise the atmosphere will become unstable to Rayleigh–Taylor instabilities because the potential density will be lower than the actual density at a higher altitude. Solving the above for the limiting case of $T{=}T_{\rm B}$ (i.e., instabilities occurring only at the top of the atmosphere) and ${\rm d}\ln{\left(\rho\right)}/{\rm d}\ln{\left(r\right)}{=}0$ yields the maximum luminosity at which hydrostatic equilibrium can be maintained in a purely radiative atmosphere,
\begin{equation}
    L_{0}= \frac{64 \pi \sigma G M_{\rm n} T_{\rm B}^{4}}{3 \bar{\kappa} P_{\rm B}},
\end{equation}
so that
\begin{equation}
    \left.\frac{{\rm d}\ln{\left(\rho\right)}}{{\rm d}\ln{\left(r\right)}}\right|_{\rm rad}= \frac{GM_{\rm n} \bar{\mu}}{4 k_{\rm B}T r} \left[\left(\frac{T_{\rm B}}{T}\right)^{4}\left(\frac{4L}{L_{0}} - 1\right) - 3\right].
\label{eq:drho_final}
\end{equation}
As we will demonstrate later, an atmosphere will start developing a convective layer before the internal luminosity $L$ reaches $L_{0}$. Thus, $L_{0}$ does not represent an absolute upper limit for atmospheric stability, but rather the theoretical maximum luminosity for an ideal purely radiative atmosphere. In reality, convection allows atmospheres to remain stable at luminosities exceeding $L_{0}$. Our equation for the temperature gradient of a radiative atmosphere is therefore,
\begin{equation}
    \left.\frac{{\rm d}\ln{\left(T\right)}}{{\rm d}\ln{\left(r\right)}}\right|_{\rm rad}= -\frac{GM_{\rm n} \bar{\mu}}{4 k_{\rm B} T r} \left[\left(\frac{T_{\rm B}}{T}\right)^{4}\left(\frac{4L}{L_{0}} - 1\right) + 1\right].
\label{eq:dT_final}
\end{equation}
To evaluate whether convection takes place, we consider the Schwarzschild criterion, \\$\left[{\rm d}\ln{(T)}/{\rm d}\ln{(P)}\right]_{\rm rad}{>}\left[{\rm d}\ln{T}/{\rm d}\ln{P}\right]_{\rm ad}$, where the radiative gradient is found to be
\begin{equation}
    \left.\frac{{\rm d}\ln{\left(T\right)}}{{\rm d}\ln{\left(P\right)}}\right|_{\rm rad} = \frac{1}{4} \left[\left(\frac{T_{\rm B}}{T}\right)^{4}\left(\frac{4L}{L_{0}} - 1\right) + 1\right],
\end{equation}
by combining Equations~\ref{eq:hydrostatic} and \ref{eq:dT_final}; whereas the adiabatic gradient is $\left[{\rm d}\ln{(T)}/{\rm d}\ln{(P)}\right]_{\rm ad}{=}(\gamma{-}1)/\gamma$. We thus see that convection occurs only for,
\begin{equation}
    \frac{T}{T_{\rm B}} \begin{cases}
    < \left[\frac{\gamma}{3\gamma - 4} \left(\frac{4L}{L_{0}} - 1\right)\right]^{\frac{1}{4}}, & \text{if } \gamma {>} \frac{4}{3} \\
    > \left[\frac{\gamma}{3\gamma - 4} \left(\frac{4L}{L_{0}} - 1\right)\right]^{\frac{1}{4}}, & \text{otherwise}
    \end{cases}.
\label{eq:convection}
\end{equation}
Moreover, we observe that the lower regions of the atmosphere are radiative and the upper regions are convective for $\gamma{>}4/3$, whereas the lower regions are convective and the upper regions are radiative for $\gamma{<}4/3$. The latter condition is satisfied for a hydrogen atmosphere only when temperatures are sufficiently high for gas to be partially ionized and $\gamma{\approx}1.1{-}1.3$ \citep[$T{>}10^{4}~{\rm K}$][]{Capitelli2008,Capitelli2009}. Such high temperatures can be achieved only through shock heating from giant impacts \citep{Benz1989,Cameron1997,Canup2008,Nakajima2015,Lock2018} and are thus not relevant to this phase of planet formation though planetesimal accretion.

\begin{figure}
    \centering
    \includegraphics[width=1\linewidth]{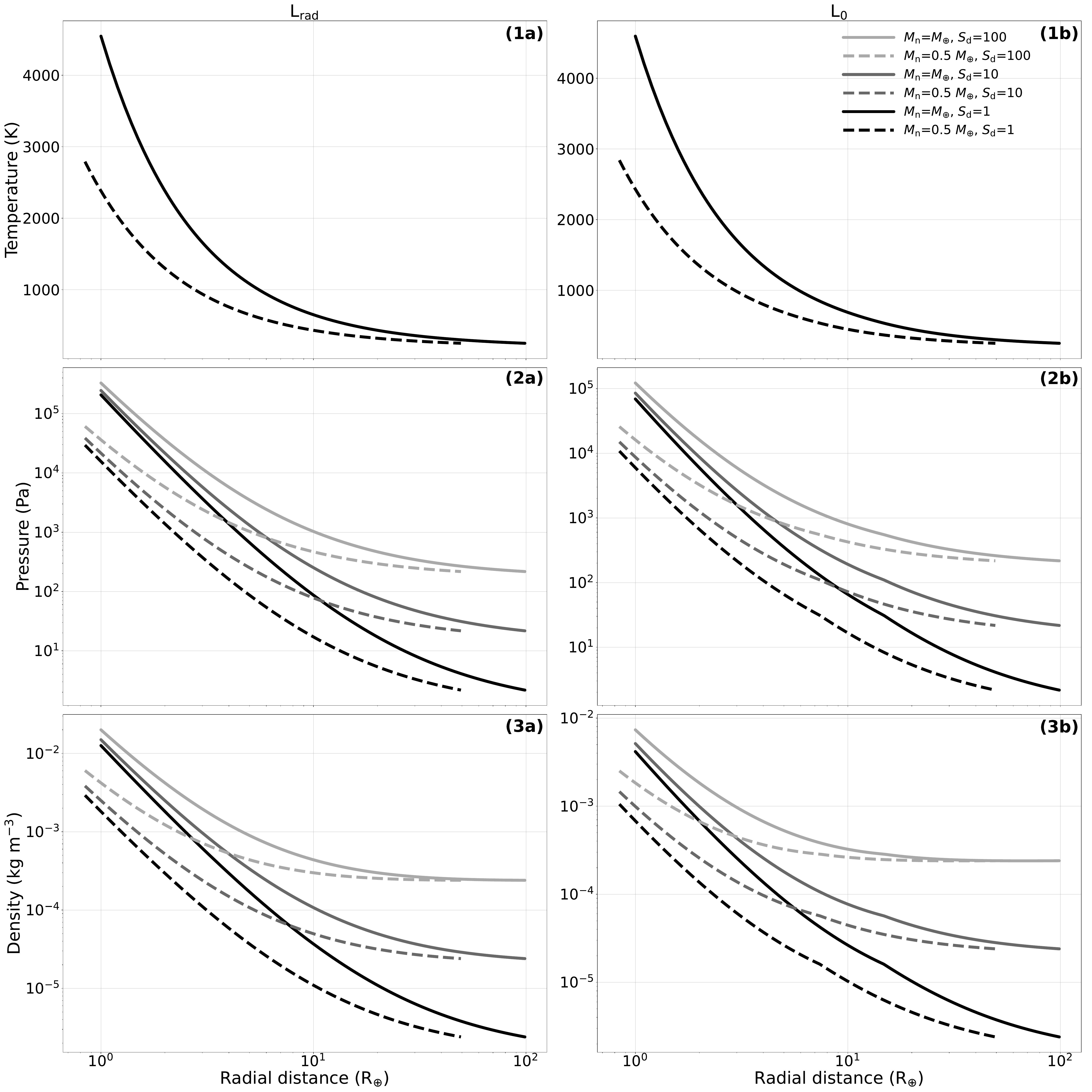}
    \caption{The temperature, density, and pressure profiles as a function of radial distance for $M_{\rm n}{=}M_{\oplus}$ (solid line) and $0.5~M_{\oplus}$ (dashed line) with $S_{\rm d}{=}1$ (black), 10 (gray), and 100 (light gray), and internal luminosities of $L_{\rm rad}$ and $L_{0}$. The minimum radius of $r{=}0.85~R_{\oplus}$ corresponds to a $M_{\rm n}{=}0.5~M_{\oplus}$ proto-Earth (see text below Equation~\ref{eq:R_B}).}
    \label{fig:scalings}
\end{figure}

\begin{figure}
    \centering
    \includegraphics[width=0.65\linewidth]{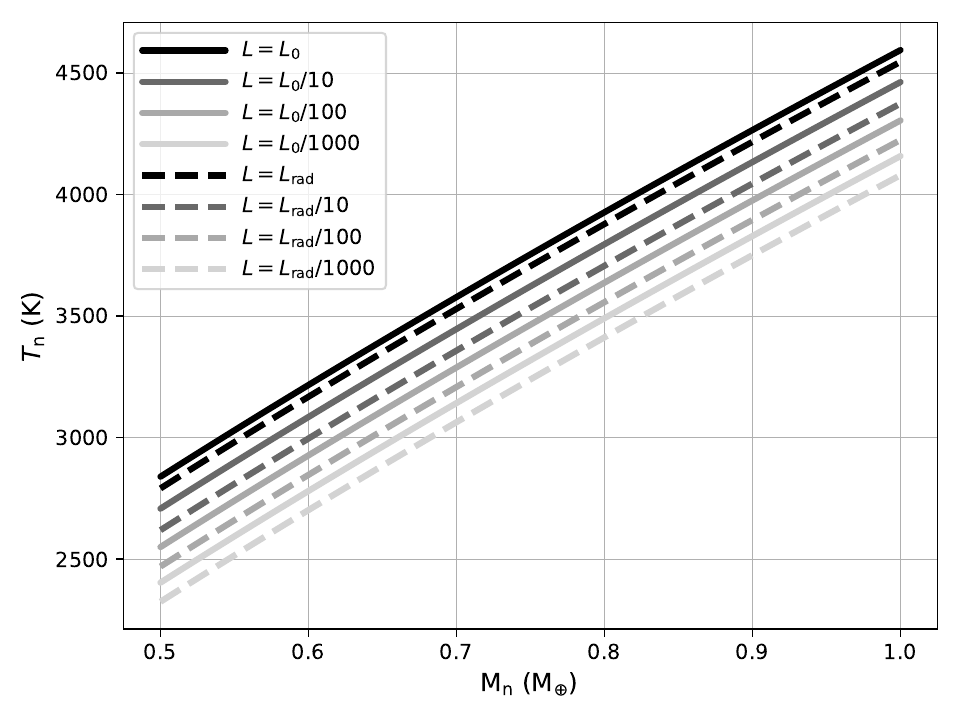}
    \caption{The surface temperature ($T_{\rm n}$) of a proto-Earth as a function of the nuclear mass ($M_{\rm n}$) for internal luminosities $L_{0}$, $L_{0}/10$, $L_{0}/100$, $L_{0}/1000$, $L_{\rm rad}$, $L_{\rm rad}/10$, $L_{\rm rad}/100$, and $L_{\rm rad}/1000$. The sensitivity to $S_{\rm d}$ is small (i.e., ${\rm d}T_{\rm n}/{\rm d}S_{\rm d}{\approx}2~{\rm K}$).}
    \label{fig:T_n}
\end{figure}

The atmosphere is fully radiative when $L{<}L_{\rm rad}$, where $L_{\rm rad}/L_{0}{=}(\gamma{-}1)/\gamma$ (Equation \ref{eq:convection}) and $L_{\rm rad}{<}L_{0}$. As explained earlier, luminosities higher than $L_0$ are possible because convective instabilities can initiate in the upper atmospheric layers. To determine $L_{\rm conv}$, the minimum luminosity for a fully convective atmosphere, we start with the convective temperature gradient,
\begin{equation}
    \left.\frac{{\rm d}\ln{\left(T\right)}}{{\rm d}\ln{\left(r\right)}}\right|_{\rm ad} = -\frac{\gamma -1}{\gamma}\frac{GM_{\rm n} \bar{\mu}}{k_{\rm B} T r},
\label{eq:dT_final2}
\end{equation}
and integrate it from the Bondi radius downward to obtain:
\begin{equation}
    T(r) = T_{\rm B}\left(\frac{3-\gamma}{2}+\frac{\gamma-1}{2}\frac{R_{\rm B}}{r}\right).
\label{eq:T_adi}
\end{equation}
Combining Equation~\ref{eq:T_adi} for $r{=}R_{\rm n}$ with Equation~\ref{eq:convection} to solve for luminosity, we get:
\begin{equation}
    \frac{L_{\rm conv}}{L_{0}} = \frac{3\gamma -4}{4 \gamma}\left(\frac{3-\gamma}{2}+\frac{\gamma-1}{2}\frac{R_{\rm B}}{R_{\rm n}}\right)^{4} + \frac{1}{4}.
\label{eq:L_conv}
\end{equation}
The luminosity $L_{\rm conv}$ ranges from $1.75{\times}10^{22}~{\rm W}$ to $1.75{\times}10^{20}~{\rm W}$ for $M_{\rm n}{=}M_{\oplus}$ and $S_{\rm d}{=}1$ to 100, and from $1.21{\times}10^{21}~{\rm W}$ to $1.21{\times}10^{19}~{\rm W}$ for $M_{\rm n}{=}0.5~M_{\oplus}$ over the same $S_{\rm d}$ range. After becoming fully convecting, a boundary layer will form at the interface between the convecting atmosphere and the magma ocean below \citep[i.e., classical scaling for Rayleigh–Bénard convection;][]{Iyer2020}. Energy is transferred through conduction and radiation, which are both Fickian, that is, they have the functional form $F{\propto}{\rm d}T/{\rm d}r$ in steady state. The total heat flow is therefore given by,
\begin{equation}
    F = \left(K_{\rm c}+ K_{\rm r}\right)\frac{{\rm d}T}{{\rm d}r},
\end{equation}
where $K_{\rm c}$ and $K_{\rm r}$ are the thermal and radiative conductivity coefficients,
\begin{equation}
    K_{\rm c} = \frac{75}{64d^{2}} \frac{k_{\rm B}}{\mu} \left(\frac{\mu k_{\rm B}T}{\pi}\right)^{\frac{1}{2}}
\end{equation}
and
\begin{equation}
    K_{\rm r} = \frac{16 \sigma T^{3}}{3 \rho \bar{\kappa}},
\label{eq:radiative_cond}
\end{equation}
respectively. For standard values of $\rho{\sim}1~{\rm m^{-3}}$, $\bar{\kappa}{\sim}1~{\rm m^{2}~kg^{-1}}$, $d{\sim}2.4{\times}10^{-10}~{\rm m}$, and $\mu{\sim}2.3~{\rm amu}$, we obtain $K_{\rm r}{>}K_{\rm c}$ for $T{>}60~{\rm K}$. In other words, for the high surface temperatures we are considering, the bottom atmospheric boundary layer should be modeled with the radiative equation and not the conductive one. However, the magma ocean boundary layer, as discussed later, should still follow the standard Rayleigh–Bénard scaling because the high $\bar{\kappa}$ value for magma ensures that conduction dominates over radiative diffusion \citep{Faber2002,Faber2020}. We thus define the Advective-Radiative (Ar) number, which compares the timescales for convective heat transport to the timescale for radiative heat transport. Because the Ar and Ra numbers are derived in the same manner but with different thermal diffusivity values (i.e., $k_{\rm c}$ vs. $k_{\rm r}$), we may use the same functional form:
\begin{equation}
    {\rm Ar} = \frac{\rho g \alpha \Delta T \Delta R^{3}}{\eta k_{\rm r}},
\end{equation}
where $\rho$, $g$, $\alpha$, $\Delta T$, $\Delta R^{3}$, $\eta$, and $k_{\rm r}$ are the density, gravitational acceleration, thermal expansion coefficient, temperature contrast, thickness, viscosity, and radiative diffusivity at the atmosphere-magma interface. For radiative diffusion, $k_{\rm r}{=}lv_{\lambda}/3$, where $l$ is the mean free path of a photon, $v_{\lambda}$ is the speed of light, and the three accounts for the three dimensions of space. The mean free path is $l{=}(\bar{\kappa} \rho)^{-1}$ where $\bar{\kappa}$ is the wavelength-averaged mean opacity and $\rho$ is the density. Combining together, we obtain
\begin{equation}
    {\rm Ar} = \frac{3\bar{\kappa} \rho^{2} g \alpha \Delta T \Delta R^{3}}{\eta v_{\lambda}},
\label{eq:Ar_number}
\end{equation}
where the critical $\rm Ar_{\rm c}$ value is approximately $10^{3}$ (i.e., the critical Rayleigh number). Equation~\ref{eq:Ar_number} can be solved with the equation for radiative heat transport, $F{=}K_{\rm r}\Delta T/\Delta R$, for the temperature contrast,
\begin{equation}
    \Delta T = \left(\frac{9 {\rm Ar_{\rm c}} v_{\lambda}}{4096 \sigma^{3}}\right)^{\frac{1}{4}} \left(\frac{\rho \eta F^{3} \bar{\kappa}^{2}}{g \alpha T^{9}} \right)^{\frac{1}{4}}.
\label{eq:dT}
\end{equation}
Equation~\ref{eq:dT} applies only when the entire atmosphere is convective and an atmospheric boundary layer forms at the atmosphere-nuclear surface interface, which occurs when $L{\geq}L_{\rm conv}$. However, as shown earlier, luminosity $L_{\rm conv}$ is very high, so it is unlikely to occur for planets before the giant impact phase. Figure~\ref{fig:scalings} shows the temperature, pressure, and density profiles for $M_{\rm n}{=}0.5~M_{\oplus}$ to $M_{\oplus}$, $L{=}L_{\rm rad}$, $L_{\rm 0}$, and $T_{\rm B}{=}250~{\rm K}$.
We can also evaluate the surface temperature of the nucleus for different thermodynamic conditions (Figure~\ref{fig:T_n}). 

\begin{figure}[H]
    \centering
    \includegraphics[width=\linewidth]{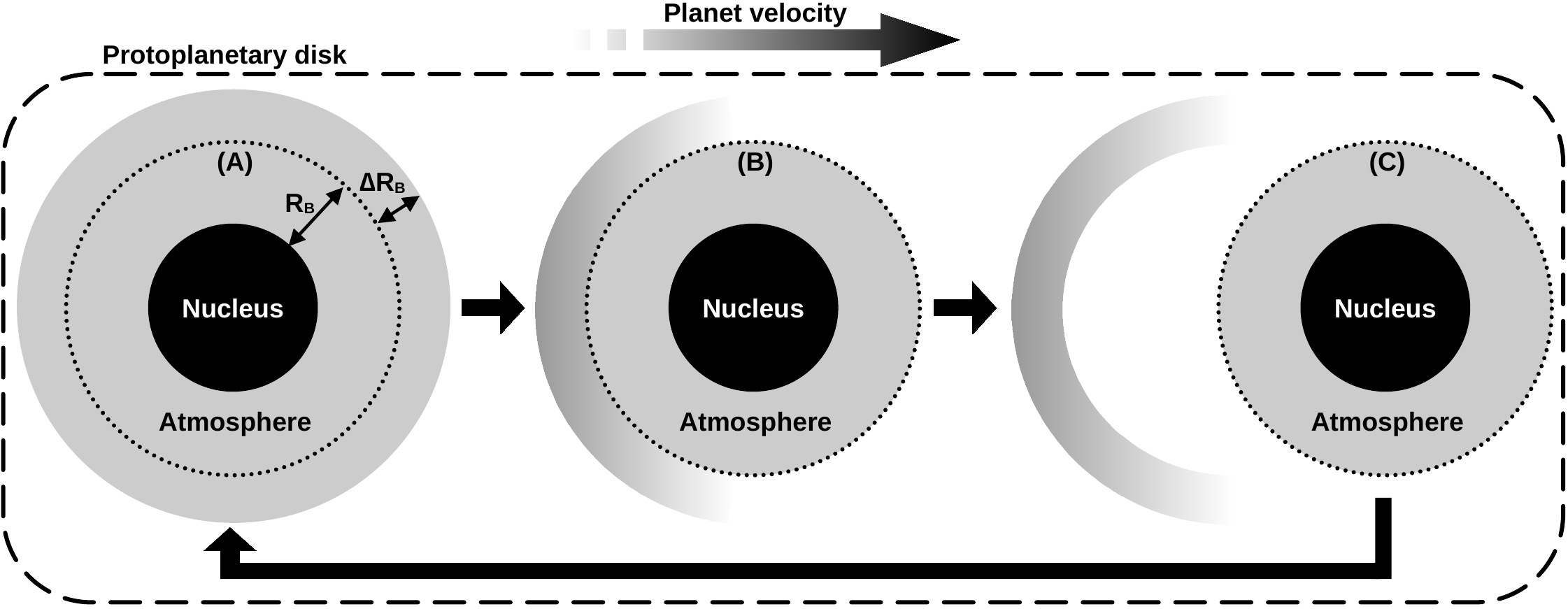}
    \caption{Schematic diagram showing the three stages of cooling from Bondi radius delaminations. In (a), the planet forms a Bondi radius boundary layer of thickness $\Delta R_{\rm B}$ with a temperature contrast $\Delta T_{\rm B}$ across it. In (b) and (c), the movement of the planet through the protoplanetary disk causes the boundary layer to become unstable and delaminate. Diagram not to scale.}
    \label{fig:delamination}
\end{figure}

The question now arises as to how a growing Earth can cool with a protoplanetary disk surrounding it. Indeed, transferring energy across the protoplanetary disk through radiative transport is highly inefficient because of its large optical depth (Equation~\ref{eq:tau_B}). Through hydrodynamic simulations, \citet{Ormel2015b} show that gas at high altitudes within the Bondi sphere is continually recycled, enabling an efficient energy exchange between the planet and protoplanetary disk. Building on this picture, we propose that cooling occurs through Bondi radius delaminations, that is, a boundary layer will form at the Bondi radius separating the atmosphere and the protoplanetary disk. This boundary layer is exposed to a drag force by the motion of a proto-Earth through the protoplanetary disk (Figure~\ref{fig:delamination}). The details of the viscous interaction between the Bondi radius boundary layer and the protoplanetary disk are not important because the Bondi radius is defined as the location where gravitational potential energy equals internal energy, so any non-zero drag force would be sufficient to cause delamination. 

The energy loss per unit area from a delamination can be estimated as $\Delta e_{\rm B}{=}\varepsilon \rho_{\rm B} \Delta R_{\rm B} c_{P} \Delta T_{\rm B}$, where $\varepsilon$ is an efficiency factor, $\rho_{\rm B}$ is the Bondi radius density, $\Delta R_{\rm B}$ is the Bondi radius boundary layer thickness, $c_{P}$ is the specific heat, and $\Delta T_{\rm B}{=}T_{\rm B}{-}T_{\rm d}\left(a_{\rm p}\right)$ is the temperature contrast across the boundary layer. The average temperature across this boundary is approximately $\left(T_{\rm B}{+}T_{\rm d}\left(a_{\rm p}\right)\right)/2$, therefore, the energy loss from delaminations has to be scaled by one half and $\varepsilon{\approx}1/2$. The delamination timescale is approximately $t_{\rm B}{\approx}\Delta R_{\rm B}/v_{\rm o}$, where $v_{\rm o}{\approx}3{\times}10^{4}~{\rm m~s^{-1}}$ is the orbital velocity of the planet because the boundary layer requires only a slight displacement to become gravitationally unbound. The cooling flux from delaminations is therefore,
\begin{equation}
    F_{\rm B}{\approx}\Delta e_{\rm B}{/}t_{\rm B}{=}\varepsilon \rho_{\rm B} c_{P} \Delta T_{\rm B} v_{\rm o}.
\label{eq:cooling_flux}
\end{equation}
For $\rho_{\rm B}{=}2.4{\times}10^{-6}~{\rm kg~m^{-3}}$ (Equation~\ref{eq:disk_density} with $S_{\rm d}{=}1$), $c_{P}{=}1.4{\times}10^{4}~{\rm J~kg~K^{-1}}$, $\Delta T_{\rm B}{=}1~{\rm K}$, and $v_{\rm o}{=}3{\times}10^{4}~{\rm m~s^{-1}}$, we obtain a cooling flux of $F_{\rm B}{\approx}500~{\rm W~m^{-2}}$ (or equivalently \\$F_{\rm n}{\approx}2{-}5{\times}10^{6}~{\rm W~m^{-2}}$ for $M_{\rm n}{=}0.5{-}1~M_{\oplus}$). Using a higher Bondi radius density corresponding to $S_{\rm d}{=}10{-}100$ (Equation~\ref{eq:disk_density}) increases the cooling flux by one to two orders of magnitude, respectively. Collectively, this suggests that cooling from delaminations is efficient even when $\Delta T_{\rm B}$ is small. This cooling flux drives the energy transfer from the top boundary layer of the cooling magma ocean to the atmosphere through conduction. Radiative and convective transport then moves this energy to the Bondi radius, where it is lost through delaminations. Therefore, understanding magma ocean cooling requires evaluating the properties of its top boundary layer. In the following section, we discuss our magma ocean model.

\section{Magma ocean model}
\label{sec:magma_model}

\begin{figure}[htbp]
    \centering
    \includegraphics[width=0.65\linewidth]{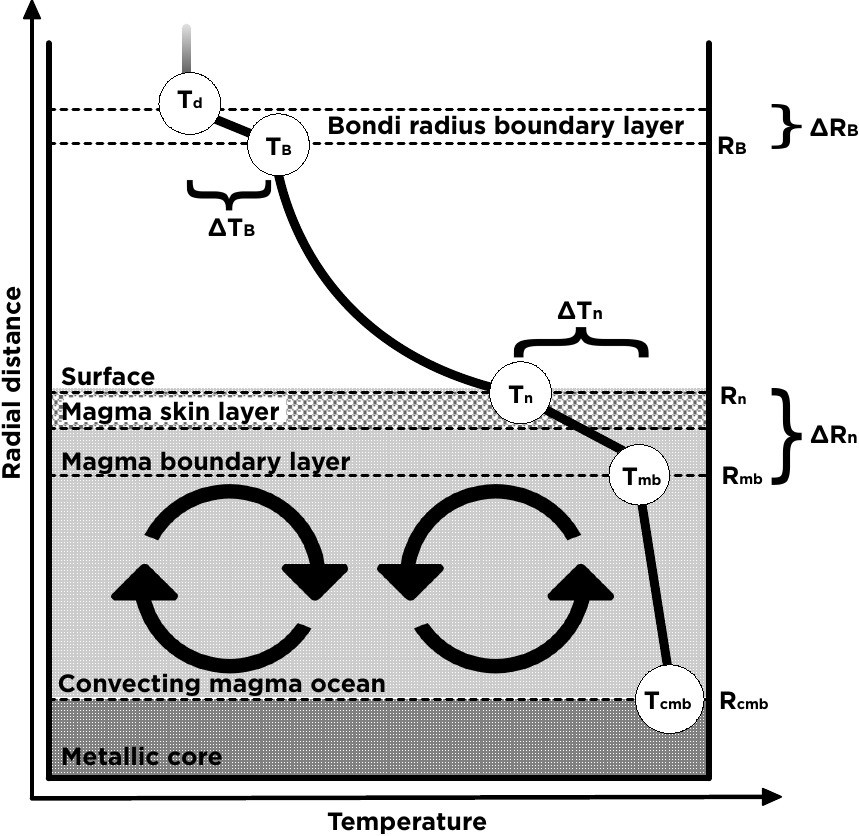}
    \caption{Schematic diagram showing the expected temperature profile of a proto-Earth within a protoplanetary disk. Temperature and distance are not to scale; $T_{\rm d}$, $T_{\rm B}$, $T_{\rm n}$, $T_{\rm mb}$, and $T_{\rm cmb}$ are the temperatures of the local protoplanetary disk gas, the Bondi radius, the surface of the nucleus, the top of the convecting magma ocean, and the metallic core-magma ocean boundary layer, respectively. Parameters $\Delta T_{\rm B}$, $\Delta R_{\rm B}$ and $\Delta T_{\rm n}$, $\Delta R_{\rm n}$ are the Bondi radius and magma ocean temperature contrasts and boundary layer thicknesses. The magma skin layer depicts the region of the magma ocean immediately in contact with the atmosphere, which has a different composition to the rest of the magma ocean because of the oxidation reactions it experiences with the atmosphere above (section~\ref{sec:dissolution}). The parameter $R$ with the relevant subscript marks the radius at which each temperature is deﬁned.}
    \label{fig:atm_profile}
\end{figure}
Our atmosphere-magma configuration is shown in Figure~\ref{fig:atm_profile}. The magma ocean convective heat flux is derived by combining the equation for the critical Rayleigh number,
\begin{equation}
    {\rm Ra}_{\rm c} = \frac{\rho_{\rm n} g_{\rm n} \alpha_{\rm n} \Delta T_{\rm n} \Delta R_{\rm n}^{3}}{\eta_{\rm n} k_{\rm n}},
\label{eq:Rayleigh_number}
\end{equation}
with the equation for steady state conduction,
\begin{equation}
    F_{\rm n} = \frac{k_{\rm n} \rho_{\rm n} c_{P,{\rm n}}\Delta T_{\rm n}}{\Delta R_{\rm n}},
\label{eq:steady_conduction}
\end{equation}
to remove dependency on the boundary layer thickness $\Delta R_{\rm n}$. This yields,
\begin{equation}
    F_{\rm n} = \left(\frac{\rho_{\rm n}^{4}c_{P,{\rm n}}^{3}g_{\rm n} \alpha_{\rm n} k_{\rm n}^{2} \Delta T^{4}_{\rm n}}{{\rm Ra}_{\rm c} \eta_{\rm n}}\right)^{\frac{1}{3}},
\label{eq:F_magma}
\end{equation}
where ${\rm Ra}_{\rm cr}{\approx}1000$, $\rho_{\rm n}$ is the density \citep[data from][parameterized by \citealt{Korenaga2023}]{Miyazaki2019b}, 
\begin{equation}
    \rho_{\rm n} = 2870-0.082T+162P^{0.58},
\label{eq:rho}
\end{equation}
$c_{P,{\rm n}}$ is the specific heat \citep{Korenaga2023}, 
\begin{equation}
    c_{P,{\rm n}}=627+0.411T-0.211P,
\label{eq:c_p}
\end{equation}
$\alpha_{\rm n}$ is the volumetric thermal expansion coefficient \citep{Korenaga2023},
\begin{equation}
    \alpha_{\rm n} = 3.622{\times}10^{-5}\exp{\left({-}2.377{\times}10^{-5}T{-}0.0106P\right)} ,
\label{eq:alpha}
\end{equation}
$\eta_{\rm n}$ is the dynamic viscosity \citep{Dingwell2004},
\begin{equation}
\eta_{\rm n}= 4.898{\times}10^{-5}\exp{\left(\frac{8526}{T-761.7}\right)},
\label{eq:eta}
\end{equation}
$k_{\rm n}$ is the thermal diffusivity \citep[average of basalt values;][]{Freitas2021},
\begin{equation}
    k_{\rm n}=8.8{\times}10^{-5}T^{-1.06}{+}8.0{\times}10^{-11}T,
\label{eq:kappa}
\end{equation}
and $\Delta T_{\rm n}{=}T_{\rm mb}{-}T_{\rm n}$ is the temperature contrast, $T_{\rm mb}$ is the mantle potential temperature, $T_{\rm n}$ is the surface temperature, and $g_{\rm n}$ is the gravitational acceleration at the surface of the nucleus. Temperature $T$ is in K and pressure $P$ is in GPa. The energy released from magma ocean cooling is approximately,
\begin{equation}
    \Delta E = \left(1{-}f_{\rm cmf}\right)M_{\rm n}\bar{c}_{P}\left(T_{\rm mb,i}{-}T_{\rm mb,f}\right),
\label{eq:dE}
\end{equation}
where $f_{\rm cmf}$ is the metallic core mass fraction, $\bar{c}_{P}{=}3000~{\rm J~K^{-1}~kg^{-1}}$ is the bulk average specific heat of the mantle including the latent heat of crystallization, and $T_{\rm mb,i}$ and $T_{\rm mb,f}$ are the initial and final temperatures, respectively, at the top of the convecting magma ocean. Equation~\ref{eq:dE} and the value of $\bar{c}_{P}$ are fit from simulations of the energy difference between two mantles of equal mass and composition but different temperature profiles. Magma ocean crystallization is highly pressure dependent, beginning from the bottom and progressing upward \citep{Solomatov2015,Miyazaki2019b}. Thus, the effects of the latent heat of crystallization are important even when the surface of the magma ocean is hot and fully molten (i.e., $T_{\rm n}{\sim}2000~{\rm K}$). Moreover, the thermodynamic properties of magma such as its density and specific heat vary with temperature and pressure, and it is therefore necessary to account for these dependencies when modeling magma ocean cooling. Our $\bar{c}_{P}$ value lies within the $2000{-}11{,}000~{\rm J~K^{-1}~kg^{-1}}$ range calculated by \citet{Korenaga2023} for a fully formed Earth (see its Figure~4).

The magma ocean of a cooling proto-Earth (i.e., with temperature $T_{\rm mb}$) can only cool asymptotically. However, it is still useful to derive a cooling timescale because it is informative of the cooling efficiency of a growing Earth. Combining Equation~\ref{eq:F_magma} with the bulk cooling flux of the mantle,
\begin{equation}
    \bar{F}_{\rm n} = \frac{M_{\rm n}\left(1{-}f_{\rm cmf}\right)\bar{c}_{P}}{4 \pi R_{\rm n}^{2}} \frac{{\rm d}T_{\rm mb}}{{\rm d}t},
\end{equation}
yields the \textit{e}-folding time,
\begin{equation}
    t_{e} = \left(T_{\rm mb}{-}T_{\rm n}\right)\frac{{\rm d}t}{{\rm d}T_{\rm mb}} = \frac{M_{\rm n}\left(1{-}f_{\rm c}\right)\bar{c}_{P}{\rm Ra}_{\rm cr}^{\frac{1}{3}}\eta_{\rm n}^{\frac{1}{3}}}{4 \pi R_{\rm n}^{2} \rho_{\rm n}^{\frac{4}{3}}c_{P,{\rm n}}g_{\rm n}^{\frac{1}{3}}\alpha_{\rm n}^{\frac{1}{3}} k_{\rm n}^{\frac{2}{3}} \left(T_{\rm mb}{-}T_{\rm n}\right)^{\frac{1}{3}}}.
\label{eq:time}
\end{equation}
Evaluating Equation~\ref{eq:time} shows that the \textit{e}-folding time of a proto-Earth of mass $0.5{-}1~M_{\oplus}$ is $10^{3}{-}10^{4}$ years. This suggests that cooling occurs rapidly, and the time frame during which hydrogen may dissolve and circulate within the magma ocean is short.

\section{Atmosphere-interior interactions}
\label{sec:interactions}

\subsection{Hydrogen dissolution model and chemistry}
\label{sec:dissolution}

Henry's law suggests that a surface magma ocean will dissolve hydrogen from the primordial atmosphere until it reaches chemical equilibrium \citep{Chachan2018,Kite2019,Kite2020,Schlichting2022}. Dissolved hydrogen reacts quickly with oxides in the magma to form water, with the amount of water depending on the type of oxides available and their abundances. The oxygen fugacity of the shallow part of a growing magma ocean is likely to be controlled by the iron-wustite buffer \citep{Deng2020}, so we focus on the reaction of $\rm FeO{+}2H{=}Fe{+}H_{2}O$. Moreover, because we are focused on the interface between the magma ocean and the atmosphere, and assuming an ideal gas, fugacity and partial pressure are interchangeable. The oxygen fugacity $f_{\rm O_{2}}$ is estimated using the analytic fit of \citet[][see their table 1]{Hirschmann2021}. We estimate the amount of water formed as follows. We start with the equilibrium constant $K{=}\exp\left({-}\Delta G^0_{\rm r}\left(T\right)/\left(RT\right)\right)$, where $K{=}f_{\rm H_{2}O}/\left(f_{\rm H}f_{\rm O_{2}}^{0.5}\right)$, $\Delta G^0_{\rm r}\left(T\right)$ is the Gibbs energy of reaction at temperature $T$, and $f_{\rm H}$, $f_{\rm O_{2}}$, and $f_{\rm H_{2}O}$ are the fugacities of atomic hydrogen, oxygen, and water. Thermodynamic data are from the NIST database \citep[i.e., The National Institute of Standards and Technology;][]{Linstrom1997}. For all regions of the parameter space we find that $f_{\rm H_{2}O}/(f_{\rm H_{2}}{+}f_{\rm H_{2}O}){>}0.99$, suggesting that virtually all dissolved hydrogen reacts with FeO to form water. Water formation breaks the equilibrium set by Henry's law between the hydrogen-rich atmosphere and the hydrogen-poor magma ocean so that the system attempts to restore equilibrium by dissolving more hydrogen. At the same time, formed water in the magma ocean skin layer is in disequilibrium with the water-poor atmosphere and therefore degasses. In other words, the system will reach chemical equilibrium when the amounts of hydrogen and water in the atmosphere and the magma ocean satisfy Henry's law.

We consider two scenarios: one where the enriched magma ocean skin layer does not mix and becomes stagnant, and one where mixing occurs. If there is no mixing, only the oxides within the skin layer can react with hydrogen to form water. We estimate the amount of water that can form in the magma ocean by considering the limit when all FeO in the magma ocean surface skin layer is consumed. The magma ocean boundary layer has a thermal (subscript mb) and compositional (subscript mb,z) component, which are usually not equal in size because the thermal and compositional diffusivities are different. With no mixing, however, the compositional boundary layer would continue to grow until its thickness is equal to the thickness of the thermal boundary layer. We thus assume that for the no-mixing case, both boundary layers are equal in size. The thickness of the boundary layer is found by rearranging Equation~\ref{eq:Rayleigh_number},
\begin{equation}
    \Delta R_{\rm n} = \left(\frac{{\rm Ra}_{\rm c} \eta_{\rm n} k_{\rm n}}{\rho_{\rm n} g_{\rm n} \alpha_{\rm n} \Delta T_{\rm n}}\right)^{\frac{1}{3}}.
\end{equation}
The total abundance of oxygen stored as FeO in the skin layer is,
\begin{equation}
    M_{\rm O_{2}} = \chi_{\rm FeO} \rho_{\rm n} 4 \pi R_{\rm n}^{2}\frac{\mu_{\rm O_{2}}}{2\mu_{\rm FeO}} \left(\frac{{\rm Ra}_{\rm c} \eta_{\rm n} k_{\rm n}}{\rho_{\rm n} g_{\rm n} \alpha_{\rm n} \Delta T_{\rm n}}\right)^{\frac{1}{3}},
\label{eq:MO2}
\end{equation}
where $\chi_{\rm FeO}{\approx}0.1$ is the FeO mass fraction, $\mu_{\rm FeO}$ and $\mu_{\rm O_{2}}$ are the molecular weights of FeO and $\rm O_{\rm 2}$, and the two in the denominator accounts for the two oxygen atoms that are needed to form one oxygen molecule. Equation~\ref{eq:MO2} informs us of the maximum amount of $\rm H_{2}O$ that can form in the skin layer, that is, $M_{\rm H_{2}O}{=}\left(2\mu_{\rm H_{2}O}/\mu_{\rm O_{2}}\right)M_{\rm O_{2}}$. Whereas the amount of water formation is indeed limited by the availability of oxides in the magma ocean skin layer, we further assume that this water eventually mixes within the bulk magma ocean, to allow a straightforward comparison with the mixing scenario. This leads to the following mass balance, 
\begin{equation}
    \chi_{\rm H_{2}O} \phi M_{\rm ma} + \frac{\mu_{\rm H_{2}O}}{\bar{\mu}}x_{\rm H_{2}O}M_{\rm atm} = M_{\rm H_{2}O},
\label{eq:MH2O}
\end{equation}
where the first term is the water content in the bulk magma ocean, assuming eventual mixing, the second is the water content in the atmosphere, and the third is the bulk total amount of water. Parameter $\chi_{\rm H_{2}O}$,
\begin{equation}
    \chi_{\rm H_{2}O} = 5.46{\times}10^{-4}\left(\frac{f_{\rm H_{2}O}}{10^{5}~{\rm Pa}}\right)^{0.47},
\label{eq:chi_H2O}
\end{equation}
is the $\rm H_{2}O$ mass fraction in the magma \citep[i.e., Henry's law;][]{Sossi2023}, $f_{\rm H_{2}O}$ is the basal atmospheric water fugacity in Pa, $\phi$ is the melt fraction of the magma ocean (we assume $\phi{=}1$, i.e., fully molten for simplicity), $M_{\rm ma}$ is the mass of the bulk magma ocean, $\mu_{\rm H_{2}O}$ is the molecular weight of $\rm H_{2}O$, $\bar{\mu}$ is the mean molecular weight of the atmosphere, $x_{\rm H_{2}O}$ is the atmospheric mole fraction of ${\rm H_{2}O}$, $M_{\rm atm}$ is the mass of the atmosphere,
\begin{equation}
    M_{\rm atm} = \int^{R_{\rm B}}_{R_{\rm n}} 4 \pi r^{2} \rho(r) {\rm d}r,
\label{eq:Matm}
\end{equation}
$\rho(r)$ is given by our atmospheric model, and $M_{\rm H_{2}O}$ is the total mass of water. Combining together, we obtain,
\begin{equation}
    \chi_{\rm H_{2}O}M_{\rm ma} + \frac{\mu_{\rm H_{2}O}}{\bar{\mu}}x_{\rm H_{2}O} M_{\rm atm} = \chi_{\rm FeO} \rho_{\rm n} 4 \pi R_{\rm n}^{2} \frac{\mu_{\rm H_{2}O}}{\mu_{\rm FeO}} \left(\frac{{\rm Ra}_{\rm c} \eta_{\rm n} k_{\rm n}}{\rho_{\rm n} g_{\rm n} \alpha_{\rm n} \Delta T_{\rm n}}\right)^{\frac{1}{3}}.
\end{equation}
We substitute $x_{\rm H_{2}O}$ with $\chi_{\rm H_{2}O}$ using Equation~\ref{eq:chi_H2O} with Dalton’s Law $f_{\rm H_{2}O}{=}x_{\rm H_{2}O}P_{\rm n}$ (${=}P\left(R_{\rm n}\right)$), resulting in a $\chi_{\rm H_{2}O}^{2.1}$ dependence. Approximating $\chi_{\rm H_{2}O}^{2.1}{\approx}\chi_{\rm H_{2}O}^{2}$, and solving with the quadratic formula, we get:
\begin{equation}
    \chi_{\rm H_{2}O} {\approx} \frac{-1 + \sqrt{1 + \chi_{\rm FeO}\frac{16 \pi R_{\rm n}^{2} \rho_{\rm n} M_{\rm atm}}{M_{\rm ma}^{2}}\frac{\mu_{\rm H_{2}O}^{2}}{\bar{\mu}\mu_{\rm FeO}}\left(\frac{P_{\rm n}}{8.75{\times}10^{11}}\right)^{-1}\left(\frac{{\rm Ra}_{\rm c} \eta_{\rm n} k_{\rm n}}{\rho_{\rm n} g_{\rm n} \alpha_{\rm n} \Delta T_{\rm n}}\right)^{\frac{1}{3}}}}{2\frac{\mu_{\rm H_{2}O}}{\bar{\mu}}\left(\frac{P_{\rm n}}{8.75{\times}10^{11}}\right)^{-1}\frac{M_{\rm atm}}{M_{\rm ma}}},
\label{eq:chi_no_mixing}
\end{equation}
which yields $\chi_{\rm H_{2}O}{\approx}10^{-11}{-}10^{-10}$ (i.e., ${\sim}10^{-7}$ oceans) for standard values of $M_{\rm n}$, $T_{\rm mb}$, and $L{=}L_{0}$. This low value arises because the amount of oxygen available in the magma ocean skin layer is too small to generate enough water to significantly oxidize the bulk magma ocean.  

If mixing occurs, the bulk magma ocean is exposed to atmospheric hydrogen. The available water is now determined by the atmospheric hydrogen reservoir so that
\begin{equation}
    \chi_{\rm H_{2}O}M_{\rm ma} + \frac{\mu_{\rm H_{2}O}}{\bar{\mu}}x_{\rm H_{2}O} M_{\rm atm} = \frac{\mu_{\rm H_{2}O}}{\bar{\mu}}M_{\rm atm},
\end{equation}
and the water concentration in the magma ocean now becomes
\begin{equation}
    \chi_{\rm H_{2}O} {\approx} \frac{{-}1 {+} \sqrt{1 + 4 \left(\frac{\mu_{\rm H_{2}O}}{\bar{\mu}}\right)^{2}\left(\frac{P_{\rm n}}{8.75{\times}10^{11}}\right)^{-1} \left(\frac{M_{\rm atm}}{M_{\rm ma}}\right)^{2}}}{2\frac{\mu_{\rm H_{2}O}}{\bar{\mu}}\left(\frac{P_{\rm n}}{8.75{\times}10^{11}}\right)^{-1}\frac{M_{\rm atm}}{M_{\rm ma}}},
\label{eq:chi_mixing}
\end{equation}
which yields $\chi_{\rm H_{2}O}{\approx}10^{-4}{-}10^{-3}$ (i.e., approximately one ocean) for the same input values. The total amount of dissolved water is probably less than what Equation~\ref{eq:chi_mixing} suggests because a magma ocean does not have to occupy the whole mantle. Deeper mantle regions could be crystallized, preventing water dissolution. Equations~\ref{eq:chi_no_mixing} and \ref{eq:chi_mixing} account for the simultaneous water enrichment of both the atmosphere and the magma, along with the chemical equilibrium between them. In other words, not all of the water formed through oxidation reactions stays stably in the magma ocean.

\subsection{Quantifying buoyancy and iron stability}
\label{sec:buoyancy}

As shown in the previous section, the amount of water that can dissolve within the magma ocean of a growing Earth depends strongly on whether the magma ocean skin layer mixes within the bulk magma ocean. In this section, we explore and evaluate the stability of the skin layer. After reacting with hydrogen, FeO is reduced to metallic Fe. We can estimate how quickly metallic iron is removed from the boundary layer by evaluating relevant timescales. First, we estimate the time it takes for an iron particle to nucleate and, second, the time required for it to sink through the magma ocean.

Full details of our heterogeneous nucleation analysis of iron particle formation are found in \ref{sec:Nucleation}; here we summarize our results. We define the formation time of an iron particle as $t^{\ast}_{1}{=}\left(JV_{\rm ch}\right)^{-1}$, where $V_{\rm ch}$ is the characteristic volume of magma containing enough dissolved metallic iron to form one critical-sized particle. The volumetric nucleation rate $J$ depends on the nucleation site density $N_{0}$. This number is very large: the magma–atmosphere interface provides many nucleation sites per unit volume, with turbulence increasing the effective surface area available for nucleation \citep{Stiassnie1991,Modirrousta2021}. Moreover, bubbles formed by wind-wave breaking \citep{Kerman1986} and magma outgassing \citep{Gaonac1996} are ubiquitous, and they offer additional sites for nucleation to occur. For simplicity, we assume one site per characteristic volume so that $N_0{=}1/V_{\rm ch}$. This leads to,
\begin{equation}
    t^{\ast}_{1} \sim  \frac{\langle a \rangle \exp{\left(\frac{\Delta G_{\rm c}}{k_{\rm B}T}\right)}}{D\left(n_{\rm c}-x_{\rm Fe}n_{\rm ma}\right)} \sqrt{\frac{64 \pi^{2}\varsigma^{3}k_{\rm B}T}{V_{\rm Fe}^{2}\Delta g^{4}}}.
\end{equation}
For a magma of temperature $T{=}5000~{\rm K}$ with $n_{\rm ma}{\sim}10^{28}~{\rm m^{-3}}$, $n_{\rm Fe}{\sim}10^{27}~{\rm m^{-3}}$, $D{\sim}10^{-8}~{\rm m^{2}~s^{-1}}$ (Equation~\ref{eq:D}), $\varsigma{\sim}1~{\rm N~m^{-1}}$ \citep{Klapczynski2022}, and $\chi_{\rm Fe}{\sim}0.1$, one obtains $t^{\ast}_{1}{\sim}0.1~{\rm s}$ for a wetting contact angle of $\phi{=}30^{\circ}$.

Having estimated the time required for an iron particle to nucleate, we now evaluate the second key timescale: the time it takes for the particle to sink through the magma ocean.
\begin{figure}[htbp]
    \centering
    \includegraphics[width=\linewidth]{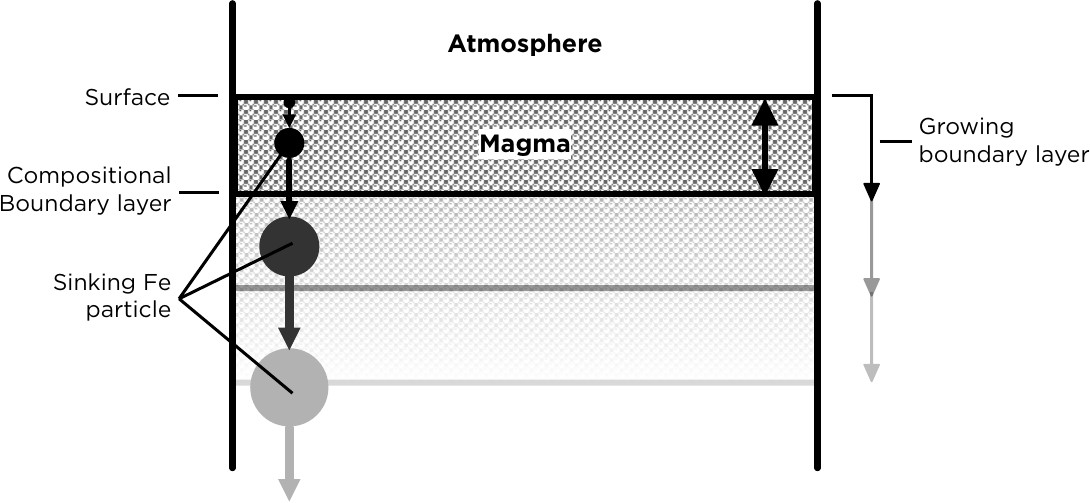}
    \caption{Schematic diagram showing the growth of the compositional boundary layer and the sinking of an iron particle. The iron particle will grow as it descends because it will acquire ambient iron.}
    \label{fig:sinking_Fe}
\end{figure}
We assume that the iron particle starts at the top of the compositional boundary layer (Figure~\ref{fig:sinking_Fe}), and it sinks vertically downward. Its characteristic velocity is given by Stokes law,
\begin{equation}
    v = \frac{2}{9} \frac{\rho_{\rm Fe}{-}\rho_{\rm ma}}{\eta} g d^{2},
\end{equation}
where $\rho_{\rm ma}$ and $\rho_{\rm Fe}$ are the densities of the magma and the metallic iron, $\eta$ is the ambient magma viscosity, $g$ is the gravitational acceleration, and $d$ is the radius of the iron particle. The radius of the particle will grow as it descends because it will acquire ambient iron, $d{\sim}\left(\chi_{\rm Fe}\rho_{\rm ma}/\rho_{\rm Fe}\right)^{1/3}\left(Dt\right)^{1/2}$, where $\chi_{\rm Fe}$ is the mass fraction of metallic iron, $D$ is the compositional diffusivity, and $t$ is the time elapsed. While the iron particle sinks, the compositional boundary layer will also grow as ${\sim}\left(Dt\right)^{1/2}$. We assume that at some distance $l^{\ast}$ and time $t^{\ast}$, the particle will overtake the growing compositional boundary layer so that
\begin{equation}
    l^{\ast} = \frac{2}{9} \frac{\rho_{\rm Fe}{-}\rho_{\rm ma}}{\eta} g D \left(\chi_{\rm Fe}\frac{\rho_{\rm ma}}{\rho_{\rm Fe}}\right)^{\frac{2}{3}} \int^{t^{\ast}}_{0}t{\rm d}t= \frac{\left(\rho_{\rm Fe}{-}\rho_{\rm ma}\right) g D}{9 \eta} \left(\chi_{\rm Fe}\frac{\rho_{\rm ma}}{\rho_{\rm Fe}}\right)^{\frac{2}{3}} t^{\ast 2}.
\end{equation}
At this time and location, the distance traveled by the sinking particle is equal to the thickness of the compositional boundary layer $l^{\ast} {=} \left(D t^{\ast}\right)^{1/2}$. Solving both equations for $t^{\ast}$ yields,
\begin{equation}
    t^{\ast}_{2} = \left[\frac{9 \eta}{\left(\rho_{\rm Fe}{-}\rho_{\rm ma}\right) g D^{\frac{1}{2}}}\right]^{\frac{2}{3}}\left(\frac{1}{\chi_{\rm Fe}}\frac{\rho_{\rm Fe}}{\rho_{\rm ma}}\right)^{\frac{4}{9}}.
\end{equation}
For standard values of $\eta{\sim}10^{-3}~{\rm Pa~s}$, $\rho_{\rm ma}{\sim}3000~{\rm kg~m^{-3}}$, $\rho_{\rm Fe}{\sim}8000~{\rm kg~m^{-3}}$, $g{\sim}10~{\rm m~s^{-2}}$, $D{\sim}10^{-8}~{\rm m^{2}~s^{-1}}$ (Equation~\ref{eq:D}), and $\chi_{\rm Fe}{\sim}0.1$ one gets $t^{\ast}_{2}{\sim}0.1~{\rm s}$. This suggests that iron will be lost quickly, making the compositional boundary layer less dense than the rest of the magma. To evaluate whether mixing occurs, the effects of both iron loss and water dissolution on buoyancy need to be quantified. Whereas it is often assumed that the magma ocean skin layer is well-mixed within the bulk magma ocean, convective mixing in the presence of compositional gradients is not guaranteed.

The effects of compositional buoyancy can be incorporated into the Rayleigh number as follows \citep[see Equation~3.1 of][]{Huppert1981},
\begin{equation}
    {\rm Ra} = \frac{\rho_{\rm n} g_{\rm n} \alpha_{\rm n} \Delta T_{\rm n} l^{3}}{\eta_{\rm n} k_{\rm n}}\left(1-\frac{k_{\rm n}}{D_{\rm n}}\frac{\beta_{\rm FeO} \Delta \chi_{\rm FeO}{+}\beta_{\rm H_{2}O} \Delta \chi_{\rm H_{2}O}}{\alpha_{\rm n} \Delta T_{\rm n}} \right),
\label{eq:Rayleigh_composition}
\end{equation}
where $l$ is the depth of the magma ocean, $D_{\rm n}$ is the compositional diffusivity \citep{Lesher1996},
\begin{equation}
    D_{\rm n}=3.73{\times}10^{-6}\exp{\left(-\frac{1.7{\times}10^{5}}{RT}\right)},
\label{eq:D}
\end{equation}
$\Delta \chi_{\rm FeO}{\approx}0.1$ and $\Delta \chi_{\rm H_{2}O}$ (Equation~\ref{eq:chi_H2O}) are the differences in the FeO and $\rm H_{2}O$ mass fractions of the magma ocean skin layer and the bulk magma ocean, and $\beta_{\rm FeO}$ and $\beta_{\rm H_{2}O}$ are their volumetric compositional expansion coefficients,
\begin{equation}
    \beta = \frac{1}{V}\frac{{\rm d}V}{{\rm d}\chi},
\end{equation}
which can be estimated from their partial molar volumes $V_{\rm m}$ as follows. For FeO \citep[in basaltic melts;][]{Guo2014}, we first we express the partial molar volume in units of $\rm m^{3}$ per kg, and we denote this parameter $V^{\ast}_{\rm m}{\approx}5.14{\times}10^{-8}T{+}9.0{\times}10^{-5}~{\rm m^{3}~kg^{-1}}$. Then, we set ${\rm d} V/{\rm d} \chi{\approx}\Delta V/\Delta \chi_{\rm FeO}{=}V^{\ast}_{\rm m} M_{\rm ma}$, where $M_{\rm ma}$ is the total mass of magma in the enriched skin layer. Combining together, we obtain $\beta_{\rm FeO}{\approx}\rho_{\rm ma}V^{\ast}_{\rm m}{\approx}1.47{\times}10^{-4}T{+}0.26$. Performing a similar calculation for $\rm H_{2}O$ yields $\beta_{\rm H_{2}O}{\approx}2.53{\times}10^{-3}T{+}0.57$ \citep[average of various silicate melts;][]{Sakamaki2017}.

\begin{figure}[htbp]
    \centering
    \includegraphics[width=\linewidth]{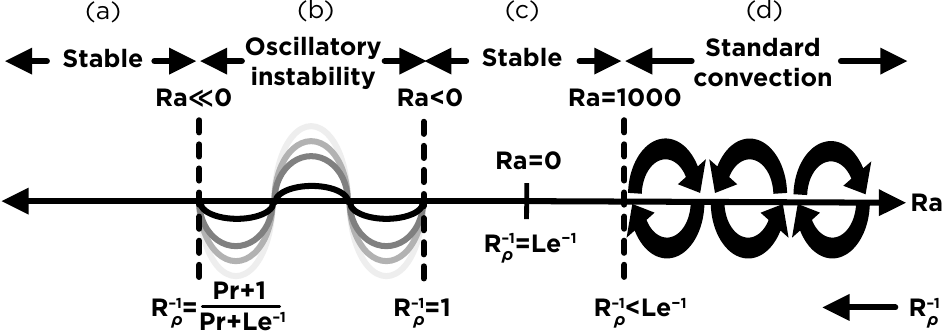}
    \caption{Illustration of the instability regimes as a function of the Rayleigh number (Equation~\ref{eq:Rayleigh_composition}) and the inverse density ratio, with $\rm Pr$ and $\rm Le$ being the Prandtl and Lewis numbers. The regimes are: (a) Stable because ${\rm Ra}{\lesssim}1000$ and $R_{\rho}^{-1}{>}R_{\rm cri}$, (b) Oscillatory instability because ${\rm Ra}{\lesssim}1000$ and $1{<}R_{\rho}^{-1}{<}R_{\rm cri}$, (c) Stable because ${\rm Ra}{\lesssim}1000$ and ${\rm Le}^{-1}{<}R_{\rho}^{-1}{<}1$ and (d) standard convection because ${\rm Ra}{\gtrsim}1000$. Rayleigh number increases from left to right whereas the inverse density ratio increases from right to left.}
    \label{fig:ddc}
\end{figure}

For clarity, we define the inverse density number and the Lewis number as $R_{\rho}^{-1}{\equiv}\left(\Sigma_{\rm i}\beta_{\rm i} \Delta \chi_{\rm i} \right)/\left(\alpha \Delta T\right)$ and ${\rm Le}{\equiv}k/D$, with $\Delta \chi$ and $\Delta T$ being the compositional and temperature contrasts, respectively. From Equation~\ref{eq:Rayleigh_composition}, we see that the Rayleigh number becomes negative when ${R_{\rho}^{-1}}{>}1/{\rm Le}$, suggesting stability against convection. When $R_{\rho}^{-1}{>}1$, the destabilizing effects of temperature are already smaller than the stabilizing effects of composition, but a different type of instability known as oscillatory double-diffusive instability may still occur \citep{Walin1964,Shirtcliffe1967,Garaud2018,Garaud2020}. Consider a section of the iron-depleted and water-rich buoyant layer (Figure~\ref{fig:atm_profile}) that is displaced vertically downward. In the absence of diffusion, this parcel would be more buoyant than its surroundings and rise again. However, if the thermal diffusivity is greater than the compositional diffusivity \citep[which is the case for magma;][]{Lesher1996,Freitas2021}, the parcel will absorb heat from its surroundings while remaining iron-depleted and water-rich before rising again. This additional heating makes the parcel more buoyant than before, thus overshooting its previous initial location, and leading to an oscillation of growing amplitude. The criterion for this mechanism is $1{<}R_{\rho}^{-1}{<}R_{\rm cri}$ \citep[see Figure~\ref{fig:ddc} and][]{Garaud2018}, where
\begin{equation}
    R_{\rm cri} = \frac{{\rm Pr}+1}{\rm Pr+{\rm Le}^{-1}},
\label{eq:R_critical}
\end{equation}
and $\rm Pr$ is the Prandtl number \citep{Walin1964,Shirtcliffe1967}.
Whereas oscillatory double-diffusive instability is less efficient than standard thermal convection \citep{Shirtcliffe1973,Linden1978}, it is still more effective in mixing than diffusion alone. 
\begin{figure}[htbp]
    \centering
    \includegraphics[width=\linewidth]{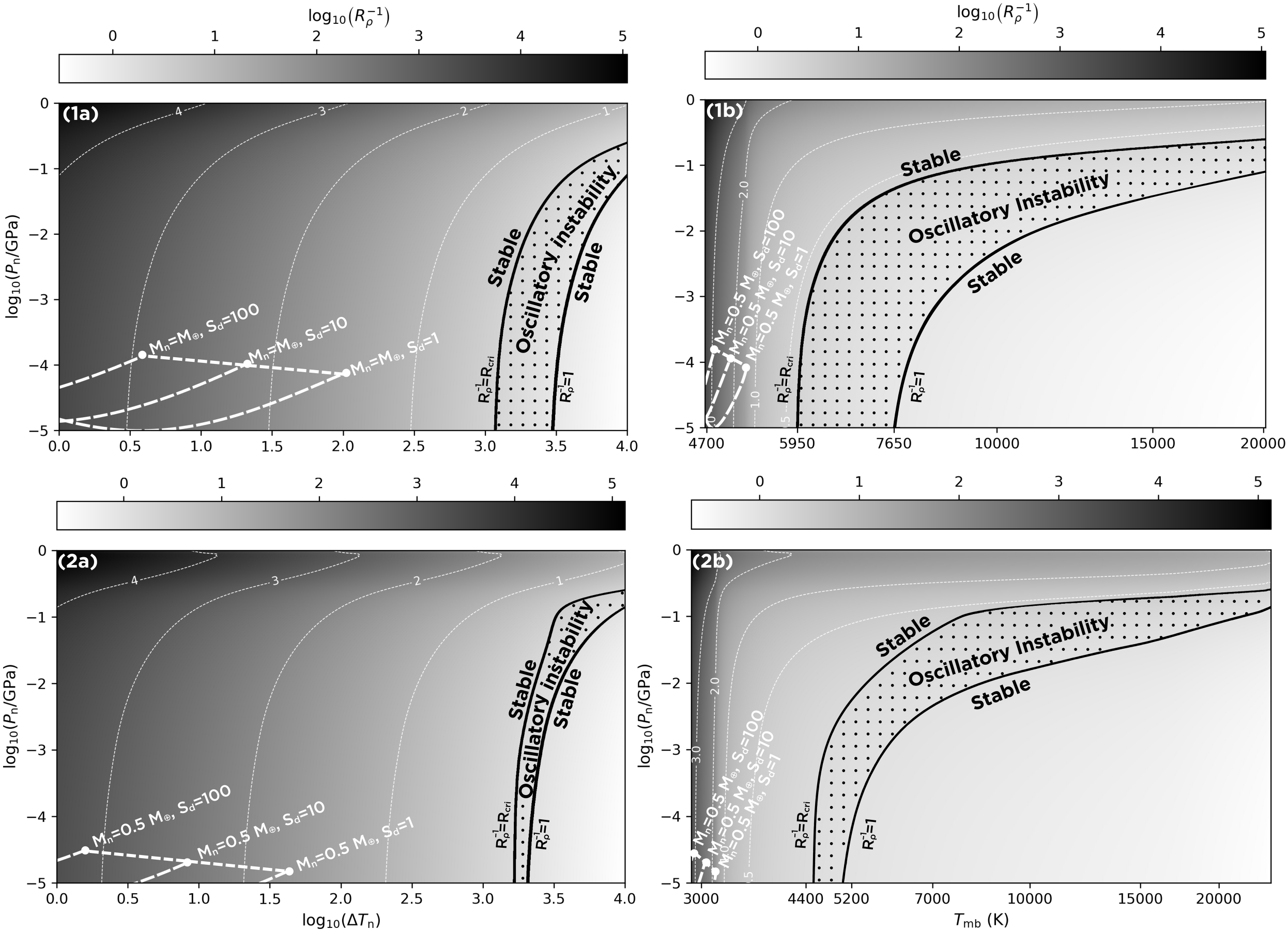}
    \caption{Inverse density ratio for a (1) $M_{\rm n} {=} M_{\oplus}$ and (2) $M_{\rm n} {=} 0.5~M_{\oplus}$ proto-Earth as a function of (a) the surface pressure and the temperature contrast across the top magma ocean boundary layer and (b) the surface pressure and the temperature of the convecting magma ocean. The thick white dashed lines indicate the domain of a proto-Earth for a given $\Delta T_{\rm n}$ when $S_{\rm d}{=}1$, 10, and $100$, with cut-offs at $L{=}L_{\rm conv}$ (Equation~\ref{eq:L_conv}). Regions hatched by the black dots mark domains in which mixing can take place. Convection is possible only when ${R_{\rho}^{-1}}{<}1/{\rm Le}$, which lies outside the domain of the graph (i.e., $\Delta T_{\rm n}{>}10^{4}~{\rm K}$ and $T_{\rm mb}{\gtrsim}20{,}000~{\rm K}$). Oscillatory double-diffusive instability occurs between the lines ${R_{\rho}^{-1}}{=}R_{\rm cri}$ and ${R_{\rho}^{-1}}{=}1$ and it requires convective stability, that is, ${R_{\rho}^{-1}}{>}1/{\rm Le}$ to operate. Neither thermochemical convection nor oscillatory double-diffusive instability occur in the $\Delta T_{\rm n}{-}T_{\rm mb}{-}P$ parameter space of a growing Earth.}
    \label{fig:ddc2}
\end{figure}
From Figure~\ref{fig:ddc2} we see that thermochemical convection and double diffusive oscillatory instability do not occur because they require unrealistically large values of $\Delta T_{\rm n}{>}10^{4}~{\rm K}$ and $\Delta T_{\rm n}{\sim}10^{3}{-}10^{3.5}~{\rm K}$ to activate. 

An alternative approach for evaluating the stability of the buoyant skin layer is to consider whether convective eddies apply sufficient shear stress to cause subduction and mixing \citep[][cf. \citealt{Martin1988,Lavorel2009}]{Sparks1984,Solomatov1993a,Solomatov1993b}. Such mechanical instability is described through the Shield's parameter \citep{Charru2004,Lajeunesse2010,Sturtz2021},
\begin{equation}
    \zeta < \frac{\sigma}{\Delta \rho g d},
\label{eq:Shields_1}
\end{equation}
where $\sigma$ is the shear stress from convective motions, $\Delta \rho$ is the density contrast between the enriched fluid and the bulk fluid, and $d$ is the characteristic length. The Shield's parameter is normally used for immiscible constituents like water and gravel \citep{Solomatov1993b}, though it likely applies also to miscible constituents because it is defined as the ratio of the shear stress to the apparent weight (i.e., weight minus buoyancy force), which apply to all mediums. In this context, the characteristic length is the boundary layer thickness $d{=}\Delta R_{\rm n}$, and the density contrast is $\Delta \rho_{\rm n} {=} \rho_{\rm n} \left(\alpha_{\rm n} \Delta T_{\rm n} {-}\beta_{\rm FeO} \Delta \chi_{\rm FeO} {-} \beta_{\rm H_{2}O} \Delta \chi_{\rm H_{2}O}\right)$. The convective shear stress is $\sigma{=}\eta_{\rm n} \bar{v}/\Delta R_{\rm n}$, with $\bar{v}$ being the characteristic velocity of convection motions. Combining all of the above we obtain,
\begin{equation}
    \zeta < \frac{\eta_{\rm n} \bar{v}}{\rho \left(\alpha_{\rm n} \Delta T_{\rm n} {-}\beta_{\rm FeO} \Delta \chi_{\rm FeO} {-} \beta_{\rm H_{2}O} \Delta \chi_{\rm H_{2}O} \right)g_{\rm n} \Delta R_{\rm n}^{2}}.
\label{eq:Shields_2}
\end{equation}
Dependency on the velocity can be removed by recognizing that at the boundary layer the advection and diffusion timescales are equal, so $\bar{v}{\approx}k_{\rm n}/\Delta R_{\rm n}$ and
\begin{equation}
    \zeta < \frac{\eta_{\rm n} k_{\rm n}}{\rho_{\rm n} g_{\rm n} \alpha_{\rm n} \Delta T_{\rm n} \Delta R_{\rm n}^{3}} \left(1 {-}\frac{\beta_{\rm FeO} \Delta \chi_{\rm FeO} {+} \beta_{\rm H_{2}O} \Delta \chi_{\rm H_{2}O}}{\alpha_{\rm n} \Delta T_{\rm n}}\right)^{-1}.
\label{eq:Shields_3}
\end{equation}
Combining with Equation~\ref{eq:Rayleigh_composition} we obtain,
\begin{equation}
    \zeta < \frac{1 {-}\frac{k_{\rm n}}{D_{\rm n}}\frac{\beta_{\rm FeO} \Delta \chi_{\rm FeO} {+} \beta_{\rm H_{2}O} \Delta \chi_{\rm H_{2}O}}{\alpha_{\rm n} \Delta T_{\rm n}}}{1 {-}\frac{\beta_{\rm FeO} \Delta \chi_{\rm FeO} {+} \beta_{\rm H_{2}O} \Delta \chi_{\rm H_{2}O}}{\alpha_{\rm n} \Delta T_{\rm n}}} = \frac{1 {-}{\rm Le}{\cdot}{\rm R}^{-1}_{\rho}}{1 {-} R^{-1}_{\rho}},
\label{eq:Shields_4}
\end{equation}
which is solved for the inverse density ratio,
\begin{equation}
    R^{-1}_{\rho} < \frac{1-\zeta}{{\rm Le}-\zeta}.
\label{eq:Shields_5}
\end{equation}
Experiments suggest that convective motions can subduct and mix a floating lid when the Shield's parameter is $\zeta{\gtrsim}0.1$ \citep{Charru2004,Sturtz2021}, which is significantly smaller than the Lewis number (${\rm Le}{\gtrsim}10$ for $T{<}5000~{\rm K}$). Equation~\ref{eq:Shields_5} thus simplifies to the previously derived relation ${R_{\rho}^{-1}}{<}1/{\rm Le}$ (Figure~\ref{fig:ddc2}). Our boundary layer analysis therefore indicates that neither thermochemical convective mixing nor oscillatory double-diffusive instabilities occur for the buoyant magma ocean skin layer because the temperature contrast is too small to trigger either mechanism.

\section{Mixing from impacts and turbulent processes}
\label{sec:mechanical mixing}

We have so far shown that a stable buoyant magma ocean skin layer will form under equilibrium conditions. One may question, however, whether disequilibrium mechanisms such as impacts and wind waves break the buoyant skin layer and allow for efficient water formation. First, we evaluate the importance of impacts.

\subsection{Mixing from impacts}

Impacts can perforate through the buoyant magma ocean skin layer and generate gas cavities that expose deeper, iron-rich regions of the magma to atmospheric hydrogen, enabling redox reactions to occur. The amount of water formation depends on the size and behavior of the impactor, such as whether it remains intact or fragments upon impact. We first consider the case where the impactor remains intact, followed by the case where fragmentation occurs.

\subsubsection{Impacts without fragmentation}

\begin{figure}[h]
    \centering
    \includegraphics[width=0.75\linewidth]{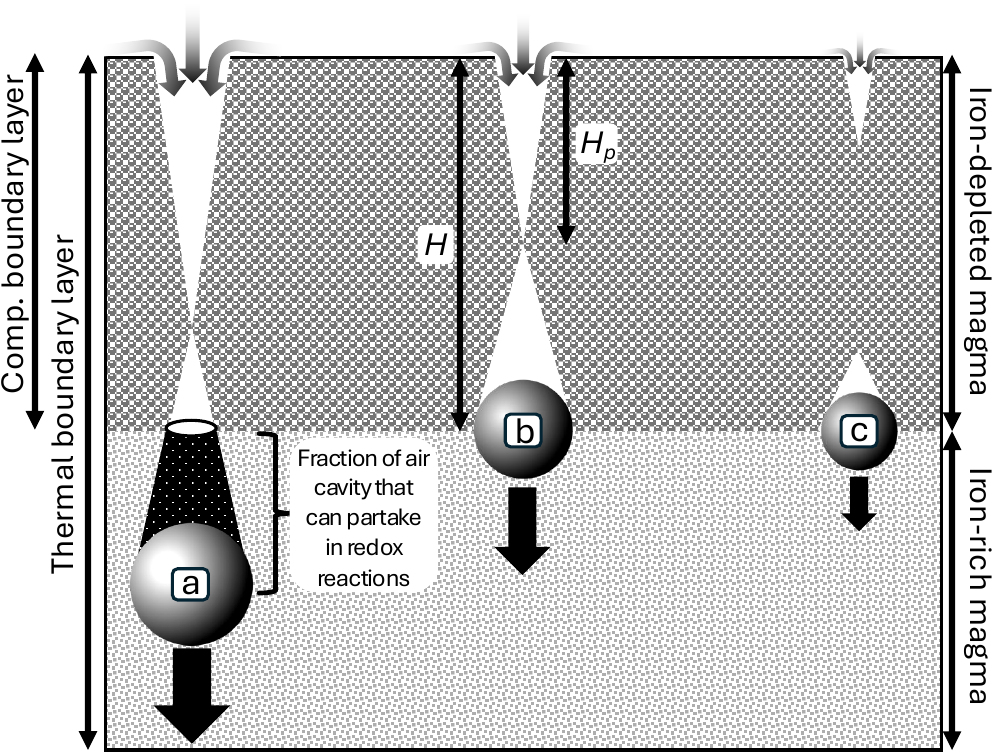}
    \caption{Schematic showing the cavity formed for three different sizes impactors. Impactor (a) is large and creates a cavity that perforates through the compositional boundary layer, (b) has the critical size for producing a cavity that equals the depth of the compositional layer, and (c) is too small and forms a cavity that collapses before reaching the iron-rich section of the magma. Here $H$ is the maximum depth of the cavity and $H_{\rm p}$ is the depth at which the cavity pinches and closes off. The black region in the lower section of cavity (a) can partake in redox reactions. The gray arrows indicate the inflowing hydrogen gas from the atmosphere and the black arrows show the direction of the sinking impactor.}
    \label{fig:ingassing}
\end{figure}

For impacts to contribute significantly to mixing, they must break the surface boundary layer repeatedly, exposing a sufficient abundance of oxides to atmospheric hydrogen for redox reactions to occur. Experiments and theory indicate that the maximum gas cavity depth, $H$, scales approximately linearly with the Froude number for low Froude numbers \citep{Duclaux2007,Yan2009} and then converges to a nearly constant value at high Froude numbers \citep{Li2020,Wang2024}. The experimental data of \citet[][black asterisks in their Figure 7a]{Wang2024} is well-fitted by the function (goodness of fit $R^{2}{=}0.97$),
\begin{equation}
    H/R=18.4-\frac{16.4}{1+\left(\frac{\rm Fr}{15.7}\right)^{4}},
\label{eq:cavity}
\end{equation}
where the Froude number is defined as ${\rm Fr}{=}u/\sqrt{2gR}$, with $u$ being the impact velocity, $g$ the gravitational acceleration, and $R$ the impactor radius. After reaching this length, pinching and collapse initiates at approximately the midway point, with the rest of the cavity imploding thereafter (Figure~\ref{fig:ingassing}). The infilling magma is sourced from local materials at similar depth levels. Thermal and chemical diffusion rapidly bring the perturbed magma back to equilibrium. The characteristic timescales for this reconstitution are $t{\sim}R^2/(2k)$ and $t{\sim}R^2/(2D)$ for thermal and compositional diffusion, corresponding to approximately one year for a $R{=}10~{\rm m}$ radius impactor and about one hundred years for a $R{=}100~{\rm m}$ radius impactor. In our analysis, the impact velocity is set by escape velocity $u{=}\sqrt{2gR_{\rm n}}$, where $R_{\rm n}$ denotes the radius of a proto-Earth. Equation~\ref{eq:cavity} assumes that the impactor remains intact through the collision process. However, if fragmentation occurs, the effective cavity depth would be considerably reduced because the impact energy would be dispersed among multiple fragments, diminishing the coherent force needed to create a deep cavity (Section~\ref{sec:fragmentation}). We apply Equation~\ref{eq:cavity}, which was originally derived from laboratory experiments of projectiles entering water. Whereas magma is three times denser than water and should produce smaller cavity depths, using this equation helps us establish an upper bound on mixing. The decelerating effects of air resistance are neglected because the impactor has large inertia.

The cavity volume is approximated as a cylinder with cross-sectional area $\pi R^{2}$ and length $H$, from which we subtract the volume occupied by the impactor (approximated as a concave spherical cap). The effective cavity volume is expressed as $V_{c}{=}\pi R^{2}\left(H{-}l\right){-}2\pi R^{3}/3$ because only the volume below the buoyant magma ocean skin layer (thickness $l$) is available for redox reactions. Gas within the buoyant skin layer does not contribute to redox reactions because there is no remaining iron oxide, and any produced gas will rise because of buoyancy, escaping the magma ocean without interacting with deeper iron-rich magma. For the impactor to perforate the buoyant magma ocean skin, the cavity length must be at least $\sqrt{2}$ times the layer’s thickness, i.e., $l{\approx}2\sqrt{Dt}$, where $D$ is the mass diffusivity and $t$ is the layer’s age. The factor of $\sqrt{2}$ arises from the average $45^{\circ}$ impact angle \citep{Robertson2021}, which increases the effective path length through the layer by a factor of $1/\cos(45^\circ){=}\sqrt{2}$. Balancing these length scales and using $M_{\rm t}{=}0.5~M_{\oplus}$, $R_{\rm n}{=}0.85~R_{\oplus}$, $D{=}10^{-8}~{\rm m^{2}~s^{-1}}$, and $t{=}1~{\rm Myr}$, yields $R_{\rm min}{=}60~{\rm m}$. Here, $R_{\rm min}$ represents the smallest impactor capable of creating a gas cavity deep enough to perforate the buoyant skin layer. Impactors smaller than 60 m create cavities that are too shallow because hydrogen only contacts the iron-depleted magma, where redox reactions cannot occur. In contrast, impactors larger than 60 m generate cavities that extend into the iron-rich magma. When these deeper cavities collapse, the inward flow of magma turbulently interacts with the hydrogen gas, enabling redox reactions to occur.

We next describe the distribution of impactors. Observations of main belt asteroids, near-Earth objects \citep{Stokes2003,Bottke2005}, and theoretical models \citep{Tanaka1996,OBrien2003} indicate that bodies in a collisional cascade exhibit a steady-state radius distribution following ${\rm d}n/{\rm d}R{\propto}R^{-7/2}$ \citep{Dohnanyi1969} or ${\rm d}n/{\rm d}M{\propto}M^{-7/4}$ when $R{\propto}M^{0.3}$ \citep{Zeng2016}. This trend holds only for objects with radii $R{\lesssim}130~{\rm km}$, suggesting larger bodies have not reached dynamic equilibrium \citep{Lissauer2013}. Protoplanetary disks dissipate rapidly \citep{Mamajek2009,Williams2011}, so it is unclear whether this equilibrium distribution is applicable. For simplicity, we assume this trend across all scales. To find the constant of proportionality for the collisional cascade, $C$, we start by calculating the total mass of impactors,
\begin{equation}
    M_{t}{=}\sum^{N}_{i} M_{i}{\approx}\int_{0}^{N} M{\,}{\rm d}n{=}\int_{0}^{N} M \left({\rm d}n/{\rm d}M\right){\,}{\rm d}M{=}C\int_{M_{\rm min}}^{M_{\rm max}} M^{-3/4}{\,}{\rm d}M,
\end{equation} 
where $M_{\rm min}$ and $M_{\rm max}$ are the smallest and largest impactors of the cascade. Evaluating this integral for $M_{\rm max}{\gg}M_{\rm min}$ and $M_{t}{\sim}M_{\rm max}$, yields $C{=}0.25M_{t}^{3/4}$.

We now estimate the cumulative volume of hydrogen drawn into the magma ocean by impactors. The total entrained volume is obtained by integrating the cavity volume over the distribution of impactors between the limits $R_{\rm min}$ and $R_{\rm max}$. For $R_{\rm min}$ we use the previously derived value of 60~m and for $R_{\rm max}$ we use $0.5~R_{\oplus}$, corresponding to a maximum impactor mass of $M_{\rm max}{=}0.1~M_{\oplus}$. We do not use a larger value because its contribution to water formation is questionable: the extreme heating of both a growing planet and its atmosphere would prevent the hydrogen envelope from remaining hydrodynamically stable, and an atmosphere is likely to be absent during the time that turbulent mechanical mixing from the impact takes place \citep{Modirrousta2023}. Of course, a new hydrogen envelope would be quickly sourced from the surrounding protoplanetary disk, but after the period of such turbulent mixing. Thus, to facilitate water formation through the efficient entrainment of atmospheric hydrogen into the magma ocean, impactors must be small enough to avoid catastrophic heating and atmospheric loss yet energetic enough to generate cavities that draw in significant volumes of hydrogen. The cumulative entrained volume is therefore given by,
\begin{equation}
    V_{t}{=}0.25{\,} M_{t}^{\frac{3}{4}} \int_{R_{\rm min}}^{R_{\rm max}} \pi R^{-\frac{3}{2}}\left\{\left[17.7-\frac{16.4}{1+\left(\frac{\rm Fr}{15.7}\right)^{4}}\right]R-2\sqrt{Dt}\right\}{\,}{\rm d}R.
\end{equation}
Evaluating this integral and multiplying by a basal atmospheric hydrogen density of $\rho{=}4{\times}10^{-3}~{\rm kg~m^{-3}}$ (Figure~\ref{fig:scalings}) yields a total hydrogen mass of $M_{H}{\approx}2{\times}10^{19}~{\rm kg}$. If all hydrogen forms water and none outgasses, this corresponds to approximately ${\sim}0.1$ Earth oceans. Even under this extreme assumption, the resulting water inventory remains too small to explain Earth's oceans. As shown in Section~\ref{sec:dissolution}, however, formed water in the magma ocean skin layer is in disequilibrium with the water-poor atmosphere and therefore degasses almost immediately. 

One might further question whether additional mixing could be driven by far-field shock waves or breaking waves generated by large impacts. If such waves could disrupt the boundary layer at large distances, they might allow hydrogen to come into contact with deeper, iron-rich magma and produce water. However, experimental and numerical studies indicate that shock-wave amplitudes rapidly decay in liquids because of geometric spreading and energy losses from frictional and viscous dissipation. \citet{Vogel1996} showed experimentally that shock pressures in water decay approximately as an inverse square law near the source and nearly inversely proportional at greater distances, with $65{-}85\%$ of the shock energy dissipating close to the epicenter. \citet{Saito2003} conducted both shock-tube experiments and hydrodynamic simulations of underwater blasts, demonstrating that shock pressures in water decrease rapidly with distance from the source. We can therefore safely conclude that far-field shocks cannot disrupt a boundary layer stabilized by compositional buoyancy. In other words, non-fragmenting impacts do not contribute substantially to water formation. In the next section, we explore the case in which fragmentation occurs.

\subsubsection{Impacts with fragmentation}
\label{sec:fragmentation}

High-velocity impacts can fragment impactors when the impact stress exceeds their material strength. The iron oxide component of these fragments may therefore dissolve into the iron-poor surface layer of the magma ocean. Efficient dissolution would increase the surface layer's iron content and density to match the underlying magma, eliminating its buoyancy and causing it to mix into the deeper magma ocean. We examine this mixing mechanism in detail below.

The impactor travels at approximately the escape velocity (${\sim}10~{\rm km~s}^{-1}$), which is nearly twice the speed of sound in the impactor. The impactor therefore travels approximately twice its diameter into the magma before internal sound waves can propagate information about fragmentation throughout the body. This ensures that most fragments become buried in the magma rather than being ejected. The impact energy is assumed to be largely consumed in the fragmentation process, preventing the formation of a deep impact cavity. The impactor is probably relatively cold, with a temperature close to the equilibrium temperature set by the star. Upon fracture, its fragments are negatively buoyant because their interiors remain significantly colder than the $3000{–}5000~{\rm K}$ magma ocean (Figure~\ref{fig:ddc2}). To determine the fate of these fragments, the timescale for their melting $t_{\rm melt}{=}r^2/(2k)$, is compared with the advection time required for them to sink through the magma. Here, $r$ and $k$ are the fragment radius (assumed spherical) and thermal diffusivity. The advection time is evaluated as $t_{\rm ad}{=}l/v$, where $v$ is the terminal Stokes velocity found by balancing the buoyancy and drag forces. This yields,
\begin{equation}
t_{\rm ad} = \frac{9 \eta l}{2 \Delta\rho g r^{2}}.
\end{equation}
For fragments to survive their transit through the buoyant magma ocean skin layer, i.e., $t_{\rm melt}{>}t_{\rm ad}$,
\begin{equation}
r > \left(\frac{9 \eta l k}{\Delta\rho g}\right)^{\frac{1}{4}}.
\end{equation}
Using characteristic values of $\eta{\sim}10^{-3}~{\rm Pa~s}$, $\Delta \rho{\sim}1000~{\rm kg~m^{-3}}$, $g{\sim}10~{\rm m~s^{-2}}$, $k{\sim}3{\times}10^{-7}~{\rm m^{2}s^{-1}}$ (Equation~\ref{eq:kappa}), and taking $l {\sim} \sqrt{2Dt} {\sim} 800~{\rm m}$ (with $D{\sim}10^{-8}~{\rm m^{2}~s^{-1}}$ from Equation~\ref{eq:D} and $t {\sim} 1~{\rm Myr}$), we find that fragments must have a radius exceeding ${\sim}4~{\rm mm}$ to survive transit through the buoyant layer. This threshold value can now be compared with the expected fragment sizes produced by an impactor undergoing fragmentation. The Grady-Kipp fragmentation theory provides a characteristic fragment size for brittle materials \citep{Grady1982,Atkinson1987,Yew1994},
\begin{equation}
    \bar{r} = \left(\frac{\sqrt{10}K_{\rm Ic}}{2\rho c_{\rm s}\dot{\varepsilon}}\right)^{\frac{2}{3}},
\end{equation}
where $K_{\rm Ic}$ is the fracture toughness, $\rho$ is the density of the impactor, $c_{\rm s}$ is its sound speed, and $\dot{\varepsilon}$ is the strain rate. For rocky impactors, we take $K_{\rm Ic} {\approx} 2{\times}10^{6}~{\rm Pa~m^{1/2}}$ \citep{Balme2004}, $\rho {\approx} 3000~{\rm kg~m^{-3}}$, and $c_{\rm s} {\approx} 4500~{\rm m~s^{-1}}$ \citep{Vanorio2002,Sekine2008}. The strain rate is approximated as $\dot{\varepsilon} {=} c_{\rm s}/(2R)$ \citep{Yew1994}, where $c_{\rm s}$ is the speed of sound and $R$ is the radius of the impactor. With these values, we obtain $\bar{r} {\sim} 1~{\rm cm}$ and ${\sim}4~{\rm cm}$ for a $10~{\rm m}$ and $100~{\rm m}$ radius impactor, respectively, which is larger than the ${\sim}4~{\rm mm}$ threshold derived above.

We also examine the Mott probability density function for the mass distribution of fragments in brittle fragmentation \citep{Grady2006},
\begin{equation}
    P(m)=\frac{1}{\bar{m}\sqrt{2m/\bar{m}}}e^{-\sqrt{2m/\bar{m}}},
\label{eq:probability}
\end{equation}
where $m$ is the fragment mass considered and $\bar{m}$ is the mean fragment mass. By construction, $\int_{0}^{\infty} P(m){\,}{\rm d}m {=} 1$ and $\int_{0}^{\infty} mP(m){\,}{\rm d}m {=} \bar{m}$. To determine whether the cumulative mass is dominated by many small fragments or by a few large ones, we examine the mass-weighted distribution $mP(m)$, which represents the contribution of fragments of mass $m$ to the total mass. Differentiating $mP(m)$ with respect to $m$, we find that it reaches a maximum at $m{=}\bar{m}/2$. This provides a natural cutoff to distinguish `small' fragments ($m{<}\bar{m}/2$) from `large' fragments ($m{>}\bar{m}/2$). By integrating the normalized distribution $(m/\bar{m})P(m)$ over these intervals, we determine that fragments with $m {<} \bar{m}/2$ contribute approximately $8\%$ of the total mass, whereas those with $m{>}\bar{m}/2$ account for approximately $92\%$. Consequently, the cumulative total mass is overwhelmingly dominated by the large fragments. As shown earlier, these larger fragments are the most likely to survive transit through the buoyant magma ocean skin layer. Therefore, most of the iron oxide component of impactors will sink without significantly altering the chemistry of the buoyant magma ocean skin layer. 

Fragmentation during atmospheric entry, which has been proposed as a mechanism for atmospheric enrichment in gas giants \citep{Valletta2019}, does not occur under the conditions considered here. For terrestrial planets with thin atmospheres (Figure~\ref{fig:scalings}), such as a proto-Earth, aerodynamic disruption of large impactors is negligible. \citet{OKeefe1982} demonstrated that for kilometer-scale bodies impacting Earth-sized planets, the atmosphere provides minimal resistance, resulting in negligible fragmentation or ablation. Nearly all of the impactor’s mass and kinetic energy reach the surface with little interaction with the atmosphere. Atmospheric breakup and chemical exchange can therefore be neglected for the impact scenarios evaluated in this study. In other words, fragmenting impacts do not contribute substantially to water formation.

\subsection{Mixing from wind waves}

In this section we examine whether wind waves can mix the buoyant magma ocean skin layer. A comprehensive wind wave model cannot be constructed because the conditions and properties of a proto-Earth are poorly constrained. Instead, we adopt a parsimonious approach that captures the fundamental physics. We model wind waves using Airy wave theory, in which the time-averaged energy density per unit horizontal area is $\langle e \rangle_{\rm ma} {=} (1/2) \rho_{\rm ma} g a_{\rm ma}^2$, where $\rho_{\rm ma}$ is the magma density, $g$ the gravitational acceleration, and $a_{\rm ma}$ the wave amplitude \citep[][equation~3.2.7;]{Phillips1977}. The wind supplies energy by applying drag force over a characteristic fetch, yielding an energy per unit area of $e_{\rm wi} {=} (1/2) C_d \rho_{\rm wi} u_{\rm wi}^2 l_{\rm wi}$, where $C_d$ is the drag coefficient, $\rho_{\rm wi}$ the wind density, $u_{\rm wi}$ the wind speed, and $l_{\rm wi}$ the fetch length. Equating the wave energy with the wind energy input yields the wave amplitude,
\begin{equation}
    a_{\rm wi} = u_{\rm wi}\sqrt{\frac{C_d \rho_{\rm wi} l_{\rm wi}}{\rho_{\rm ma} g}}.
\end{equation}
Using $u_{\rm wi} {=} 100~{\rm m~s^{-1}}$, $C_d {=} 3{\times}10^{-3}$ \citep{Curcic2020}, $\rho_{\rm wi} {=} 4\times10^{-3}~{\rm kg~m^{-3}}$ (see Figure~\ref{fig:scalings}), $l_{\rm wi} {=} 10^{6}$ m, $\rho_{\rm ma} {=} 3000~{\rm kg~m^{-3}}$, and $g {=} 10~{\rm m~s^{-2}}$, we obtain $a_{\rm wi} {=} 2~{\rm m}$. This amplitude is negligible relative to the buoyant skin layer thickness, $\sqrt{2Dt} {\sim} 800~{\rm m}$. Wind waves are therefore unable to mix the buoyant magma ocean skin layer.

\section{Results and Discussion}
\label{sec:results}

Our analysis indicates that neither oscillatory double-diffusive instability nor thermochemical convection is likely to occur in magma oceans, because a stable buoyant magma ocean skin layer forms largely because of iron loss accompanying hydrogen dissolution. These mixing processes require very high magma ocean temperatures, which is unlikely because this would require shock heating from giant impacts. A gradual planetary assembly cannot maintain such temperatures because of efficient cooling through Bondi radius delaminations (Equation~\ref{eq:time}). In other words, we find that water formation from oxidation reactions between primordial gas and oxides in magma is inefficient and unlikely to lead to ocean formation. Moreover, we show that neither mechanical mixing from fragmenting and non-fragmenting impactors nor wind-driven wave mixing is capable of disrupting the buoyant magma ocean skin layer, demonstrating that it remains dynamically stable against large-scale thermochemical mixing.

\subsection{Comparison with other works}
\label{sec:comparison}

Our results conflict with those of \citet{Ikoma2006} and \citet{Young2023}, who found significant water formation. Both studies assume that magma ocean mixing is efficient. \citet{Ikoma2006} model the interaction between a hydrogen-rich primordial atmosphere and the magma ocean, suggesting that atmospheric hydrogen can readily dissolve into the magma ocean and oxidize to form water. Their model relies on the assumption of a fully convective magma ocean with continuous mixing, allowing hydrogen to access the oxides throughout the entire magma ocean. Similarly, \citet{Young2023} employ a thermodynamic model that assumes chemical equilibrium between the hydrogen-rich atmosphere and the magma ocean below. They suggest that efficient convective turnover facilitates chemical exchange and equilibration between the magma ocean and the atmosphere over million-year timescales. We show that this assumption is not justified because oxidation reactions lead to the formation of metallic iron that sinks, making the top layer of the magma ocean buoyant and resistant to mixing.

Whereas \citet{Wu2018} also predict some water formation, they find that only ${\sim}0.06$ ocean form from oxidation reactions between the atmosphere and the magma ocean. The bulk of the water, approximately seven oceans, is instead attributed to delivery from chondritic material. To explain Earth's mantle D/H ratio of ${\sim}150{\times}10^{-6}$, they propose a two-step process. First, core formation initially increases the D/H ratio. Next, low D/H ratio hydrogen (${\sim}21{\times}10^{-6}$) from the protoplanetary disk dissolves into the mantle, lowering the ratio. However, such a two-step process may not be necessary because the D/H ratio of Earth's mantle already lies within the range of chondritic material \citep[$130{-}170{\times}10^{-6}$;][]{Wu2018}, suggesting that Earth's oceans could originate solely from chondritic material, without the need for hydrogen from the disk.

\subsection{Model limitations}
\label{sec:limitations}

Whereas our study provides useful insights into the interactions between a growing Earth and its primordial atmosphere, it is not without limitations. One limitation is that we assume a simple protoplanetary disk structure even though it is likely complex, with significant variations in temperature \citep{Estrada2016}, density \citep{Armitage2011}, and composition \citep{Henning2013}. Regarding our atmospheric model, we assume that gas is ideal and with constant opacities, overlooking more complex phase behavior \citep{Saumon1995,Militzer2013,Chabrier2019} and possible fluctuations in gas absorptivity \citep{Lenzuni1991,Freedman2008,Freedman2014}. Moreover, our chemical model is parsimonious, considering only hydrogen, water, and iron even though other species are present \citep[e.g., see primordial composition in][]{Lodders2010} and further enrichment is expected from interactions between vaporized magmatic species and atmospheric hydrogen \citep{Falco2024}. However, the exact composition of a growing Earth is not well constrained \citep{Mezger2020} because Earth shares isotopic similarities with enstatite chondrites \citep{Javoy2010,Dauphas2017} and carbonaceous chondrites such as CI chondrites \citep{Marty2012,Sarafian2014}, suggesting that it is composed of a mixture of each. Thus, to avoid the strong uncertainties of the geochemical parameter space, we focus only on interactions between atmospheric hydrogen and oxygen under the iron-wüstite buffer in magma. Furthermore, assuming constant opacity allowed us to derive 1D differential equations for the temperature (Equation~\ref{eq:dT_final}) and density (Equation~\ref{eq:drho_final}) that capture the basic physics taking place, as well as facilitating our modeling of the transition between radiative and convecting sections of the atmosphere.

\section{Conclusions}
\label{sec:conclusion}

The origin of Earth's oceans is one of the biggest questions in planetary science. Some have suggested that water could have formed through oxidation reactions between hydrogen gas in the primordial atmosphere of a growing Earth and oxides in its magma. This notion is exciting because it alludes to water and life being more prevalent in the universe than previously thought. Our analysis suggests, however, that this may not be the case because of the formation of an iron-depleted buoyant magma ocean layer that precludes further water formation. This study underscores the importance of incorporating physical reality, such as the plausibility of convective mixing in a magma ocean, into the models of chemical equilibrium during planetary formation.

\section{Data/Software Availability Statement}
This work is purely theoretical in nature and can be reproduced from the details provided in the main text. The Python scripts used for this work have been deposited at Zenodo \citep{Modirrousta2025}.

\section{Acknowledgements}

This work was sponsored by the US National Science Foundation EAR-2224727 and the US National Aeronautics and Space Administration under Cooperative Agreement {No.\,80NSSC19M0069} issued through the Science Mission Directorate. This work was also supported in part by the facilities and staff of the Yale University Faculty of Arts and Sciences High Performance Computing Center. We thank Mary-Louise Timmermans for providing important insights into double diffusive convection. We are also grateful to two anonymous reviewers for their constructive comments.

\bibliography{bibliography}{}

\appendix
\renewcommand{\theequation}{A\arabic{equation}}
\setcounter{figure}{0}
\renewcommand{\thefigure}{A\arabic{figure}}
\setcounter{subsection}{0}
\renewcommand{\thesubsection}{Appendix \Alph{subsection}}

\subsection{Derivation of Equation~\ref{eq:dT_dr_rad}}
\label{sec:radiative_diffusion}

Consider a gas in radiative equilibrium. This gas has photons scattering through it, which can be modeled using the conduction equation for a spherical system,
\begin{equation}
    \frac{L}{4 \pi r^{2}} = -K\frac{{\rm d}T}{{\rm d}r}.
\end{equation}
The conduction constant is defined as $K{=}k \rho c_{P}$, where $\rho$ is density, $k$ is thermal diffusivity, and $c_{P}$ is specific heat capacity. From kinetic theory, we know that the thermal diffusivity is $k{=}\bar{v} l /3$, where $\bar{v}$ is the average particle velocity, $l$ is the mean free path of a particle, and the factor of three accounts for scattering occurring in three dimensions. For photons, $\bar{v}{=}v_{\lambda}$ (the speed of light) and $l{=}\left(\rho \bar{\kappa}\right)^{-1}$. Combining together,
\begin{equation}
    \frac{L}{4 \pi r^{2}} = -\frac{v_{\lambda} c_{P}}{3\bar{\kappa}} \frac{{\rm d}T}{{\rm d}r}.
\label{eq:conduction_1}
\end{equation}
The specific heat is defined as $c_{P}{\equiv}\left({\rm d}Q/{\rm d}T\right)_{P}$ where $Q{=}4\sigma T^{4}{/}\left(\rho v_{\lambda}\right)$ for a radiative system, with $\sigma$ being the Stefan-Boltzmann constant. Therefore, $c_{P}{=}16 \sigma T^{3}{/}\left(\rho v_{\lambda}\right)$, which upon inserting into Equation~\ref{eq:conduction_1} and rearranging for the temperature gradient,
\begin{equation}
    \frac{{\rm d}T}{{\rm d}r} = -\frac{3 \bar{\kappa} \rho L}{64 \pi r^{2} \sigma T^{3}}.
\end{equation}
The above derivation satisfies energy conservation, and it is widely used in stellar astrophysics \citep[e.g.,][]{Ryan2010}.

\subsection{Heterogeneous nucleation of iron particles}
\label{sec:Nucleation}

Metallic iron particles can form efficiently when nucleation sites are present. In a magma ocean above the liquidus temperature, where crystals are absent, the magma-atmosphere interface itself \citep{Nepomnyashchy2006,Liang2019,Zhang2025} and the surface of gas bubbles \citep{Wohlgemuth2009, Wohlgemuth2010,Luhrmann2018,Fatemi2018} may serve as the nucleation centers needed for iron separation. From classical nucleation theory, the nucleation rate per unit volume is,
\begin{equation}
    J =  \beta N_{0} Z \exp{\left(-\frac{\Delta G_{\rm c}}{k_{\rm B}T}\right)},
\end{equation}
where $\beta$ is diffusion rate of iron atoms to the iron particle of critical size $r_{\rm c}$, $N_{0}$ is the number density of nucleation sites, $\Delta G_{\rm c}$ is the energy required to form a critical size particle, $k_{\rm B}$ is the Boltzmann constant, and $T$ is temperature. Here, $Z$ is the Zeldovich factor, which accounts for the probability of a particle near the critical size becoming stable rather than redissolving. The terms $\beta$, $r_{\rm c}$, $Z$, and $\Delta G_{\rm c}$ are all functions of the Gibbs free energy of forming a spherical particle of arbitrary size $r$,
\begin{equation}
    \Delta G(r) = \frac{4}{3}\pi r^{3} \Delta g + 4 \pi r^{2} \varsigma,
\label{eq:dG}
\end{equation}
with $\Delta g$ being the Gibbs free energy per unit volume and $\varsigma$ the surface tension. The $\Delta g$ term is given by the Gibbs free energy of mixing per unit volume,
\begin{equation}
    \Delta g = k_{\rm B}T\left[n_{\rm ma}\log{\left(x_{\rm ma}\right)}+n_{\rm Fe}\log{\left(x_{\rm Fe}\right)}\right],
\end{equation}
where $n_{\rm ma}$ and $n_{\rm Fe}$ are the number densities, and $x_{\rm ma}$ and $x_{\rm Fe}$ are the mole fractions of magma and iron, respectively. Evidently, $\Delta g$ is negative because energy is expended in collecting iron from the ambient magma to form the particle, whereas $\varsigma$ is positive because surface tension arises from the net attractive forces between metallic iron atoms at the surface of the particle. Consequently, Equation~\ref{eq:dG} has a maximum, and we label the radius at this maximum the critical radius $r_{\rm c}$. For particles with $r{<}r_{\rm c}$, growth increases the Gibbs free energy, making further growth thermodynamically unfavorable and causing the particle to shrink and disintegrate. For particles with $r{>}r_{\rm c}$, growth decreases the Gibbs free energy, favoring continued growth. Therefore, nucleation requires a particle to reach the critical size $r{=}r_{\rm c}$. Differentiating Equation~\ref{eq:dG} and setting to zero yields the critical radius $r_{\rm c}{=}-2\varsigma/\Delta g$. The energy required to form a critical size particle is found by inserting the critical radius into Equation~\ref{eq:dG}, yielding $\Delta G_{\rm c}{=}16 \pi \varsigma^{3}\Phi(\phi)/(3 \Delta g^{2})$. The correction factor,
\begin{equation}
    \Phi(\phi) = \frac{1}{4}\left(2 - 3\cos\phi + \cos^{3}\phi\right),
\end{equation}
accounts for the reduced energy barrier for nucleation. This reduction occurs at the magma-atmosphere interface or at bubble surfaces, and the magnitude of the reduction depends on the contact angle $\phi$ between the iron particle and these surfaces.

The diffusion rate of solute atoms is given by Fick's first law,
\begin{equation}
    \beta = -D\frac{{\rm d}n_{\rm Fe}}{{\rm d}r},
\label{eq:Ficks}
\end{equation}
where $D$ is the mass diffusivity coefficient of metallic iron in magma. Under the assumption of a discrete particle-magma boundary, we approximate Equation~\ref{eq:Ficks} as,
\begin{equation}
    \beta = D\frac{n_{\rm c}-x_{\rm Fe}n_{\rm ma}}{\langle a \rangle},
\end{equation}
where $n_{\rm c}$ is the number density of iron atoms in a particle of critical radius $r_{\rm c}$ and $\langle a \rangle{=}\left(3/\left(4\pi n_{\rm ma}\right)\right)^{1/3}$ is the average atomic spacing in magma.

The Zeldovich factor is defined as,
\begin{equation}
    Z = \sqrt{\frac{1}{2\pi k_{\rm B} T} \left| \frac{{\rm d}^2 \Delta G}{{\rm d}n^2} \right|_{n=n_c}},
\label{eq:zeldovich}
\end{equation}
with $\left|{\rm d}^2 \Delta G/{\rm d}n^2 \right|_{n=n_c}$ being the second derivative of Equation~\ref{eq:dG} with respect to number density, evaluated at the critical number density $n_{\rm c}$. Further evaluation yields,
\begin{equation}
    Z = \sqrt{\frac{V_{\rm Fe}^{2}\Delta g^{4}}{64 \pi^{2}\varsigma^{3}k_{\rm B}T}},
\label{eq:zeldovich_2}
\end{equation}
where $V_{\rm Fe}$ is the atomic volume of an iron atom. An evaluation of the above mentioned equations and its implications for iron nucleation in magma is found in Section~\ref{sec:buoyancy} of the main text.

\end{document}